\documentclass[manuscript,screen]{acmart}

\AtBeginDocument{%
  \providecommand\BibTeX{{%
    \normalfont B\kern-0.5em{\scshape i\kern-0.25em b}\kern-0.8em\TeX}}}

\setcopyright{acmcopyright}
\copyrightyear{2020}
\acmYear{2020}
\acmDOI{XXX} 

\usepackage{pgf}
\usepackage{subcaption}
\usepackage{amsmath}
\usepackage{multirow}
\usepackage[utf8]{inputenc}
\usepackage{listings}
\usepackage{footnote}
\makesavenoteenv{tabular}
\makesavenoteenv{table}

\usepackage{xcolor}
\usepackage{xargs}
\usepackage[colorinlistoftodos,textsize=footnotesize]{todonotes}

\newcommandx{\unsure}[2][1=]{\todo[linecolor=red,backgroundcolor=red!25,bordercolor=red,#1]{#2}}
\newcommandx{\change}[2][1=]{\todo[linecolor=blue,backgroundcolor=blue!25,bordercolor=blue,#1]{#2}}
\newcommandx{\info}[2][1=]{\todo[linecolor=OliveGreen,backgroundcolor=OliveGreen!25,bordercolor=OliveGreen,#1]{#2}}

\newcommand{\ten}[1]{\ensuremath{\mathbf{#1}}}

\usepackage{tikz}
\usetikzlibrary{arrows.meta}
\usetikzlibrary{shapes.geometric}
\usetikzlibrary{fit}

\tikzset{%
  >={Latex[width=2mm,length=2mm]},
            base/.style = {rectangle, rounded corners, draw=black,
                           minimum width=4cm, minimum height=1cm,
                           text centered, font=\sffamily},
  activityStarts/.style = {base, fill=blue!30},
       startstop/.style = {base, fill=red!30},
       activityRuns/.style = {base, fill=green!30},
       optional/.style = {base, fill=lightgray},
         process/.style = {base, minimum width=2.5cm, fill=orange!15,
           font=\ttfamily},
         decision/.style = {diamond, minimum width=3cm, text centered, draw=black, fill=green!30}
}

\newcommand{\code}[1]{\lstinline{#1}}
\newcommand{\codeblack}[1]{\lstinline[keywordstyle=\color{black}]{#1}}

\usepackage{lineno}

\usepackage{booktabs}
\usepackage{adjustbox}
\usepackage{array}
\newcolumntype{L}[1]{>{\raggedright\arraybackslash}p{#1}}

\begin{document}

\title{PySPH: a Python-based framework for smoothed particle hydrodynamics}
\author{Prabhu Ramachandran}
\authornote{Corresponding author}
\authornote{Authors in rough order of contributions to software and manuscript.}
\email{prabhu@aero.iitb.ac.in}
\orcid{0000-0001-6337-1720}
\author{Aditya Bhosale}
\email{adityapb1546@gmail.com}
\author{Kunal Puri}
\email{kunal.puri@numeca.be}
\author{Pawan Negi}
\email{pawan.n@aero.iitb.ac.in}
\author{Abhinav Muta}
\email{mutaabhinav@gmail.com}
\author{A Dinesh}
\email{adepu.dinesh.a@gmail.com}
\author{Dileep Menon}
\email{dileepsam92@gmail.com}
\author{Rahul Govind}
\email{rahulgovind517@gmail.com}
\author{Suraj Sanka}
\email{sankasuraj@gmail.com}
\author{Amal S Sebastian}
\email{amalssebastian@gmail.com}
\author{Ananyo Sen}
\email{ananyo.sen2@gmail.com}
\author{Rohan Kaushik}
\email{rohankaush@gmail.com}
\author{Anshuman Kumar}
\email{anshu266man@gmail.com}
\author{Vikas Kurapati}
\email{vikky.kurapati@gmail.com}
\author{Mrinalgouda Patil}
\email{mpcsdspa@gmail.com}
\author{Deep Tavker}
\email{tavkerdeep@gmail.com}
\author{Pankaj Pandey}
\email{pankaj86@gmail.com}
\author{Chandrashekhar Kaushik}
\email{shekhar.kaushik@gmail.com}
\author{Arkopal Dutt}
\email{arkopal.dutt@gmail.com}
\author{Arpit Agarwal}
\email{arpit.r.agarwal@gmail.com}
\affiliation{
  \institution{Indian Institute of
    Technology Bombay}
  \streetaddress{Department of Aerospace Engineering, IIT Bombay, Powai}
  \city{Mumbai}
  \postcode{400076}
}

\renewcommand{\shortauthors}{Ramachandran and Bhosale, et al.}

\begin{abstract}
  PySPH is an open-source, Python-based, framework for particle methods in
  general and Smoothed Particle Hydrodynamics (SPH) in particular. PySPH
  allows a user to define a complete SPH simulation using pure Python.
  High-performance code is generated from this high-level Python code and
  executed on either multiple cores, or on GPUs, seamlessly. It also supports
  distributed execution using MPI. PySPH supports a wide variety of SPH
  schemes and formulations. These include, incompressible and compressible
  fluid flow, elastic dynamics, rigid body dynamics, shallow water equations,
  and other problems. PySPH supports a variety of boundary conditions
  including mirror, periodic, solid wall, and inlet/outlet boundary
  conditions. The package is written to facilitate reuse and reproducibility.
  This paper discusses the overall design of PySPH and demonstrates many of
  its features. Several example results are shown to demonstrate the range of
  features that PySPH provides.
\end{abstract}



\maketitle

\section{Introduction}
\label{sec:intro}

Particle methods are a class of numerical methods where particles are used to
carry physical properties. These properties evolve as per the governing
differential equations of the problem. They may be used to simulate continuum
mechanics problems which is our present interest. Particle methods are
typically Lagrangian and meshless and this allows the method to handle
problems with free-surfaces and large deformations. The Smoothed Particle
Hydrodynamics method (SPH) was independently developed by \citet{lucy77} and
\citet{monaghan-gingold-stars-mnras-77}. The method has since developed to
simulate a wide variety of problems in continuum mechanics.

A typical meshless particle method implementation requires the following,
\begin{itemize}
\item representation of the field properties in terms of particles.
\item reconstruction of the field properties and their derivatives using the
  particle positions and properties alone.
\item identification of neighboring particles.
\item evolution of the particle properties over time.
\end{itemize}

There are various open-source software packages like
DualSphysics~\cite{crespo2015dualsphysics},
GADGET-4~\cite{springel2020simulating},
AquaGPUSph~\cite{cercos2015aquagpusph}, PHANTOM~\cite{phantom:2018},
SPlisHSPlasH~\cite{koschier_2019} among others which solve SPH problems
efficiently. A fairly exhaustive listing of open source and commercial
packages is provided in the review by \citet{Shadloo16}. In
Table~\ref{table:compare} many open source SPH packages and their features are
listed. The footnotes explain the respective columns.

\begin{table}[h!]
\centering
\begin{tabular}{llllllL{4cm}}
  \toprule
  {} & MPI/GPU~\footnote{Whether the software can be run on MPI/GPU?}
  & Language~\footnote{The programming language used for the package.}
  & NNPS~\footnote{Nearest neighbor particle search (NNPS) algorithms available.}
  & CI~\footnote{Whether Continuous integration testing is done?}
  & Docs~\footnote{Whether documentation is easy to access and covers the usage aand API?}
  & Notes~\footnote{The area of application of the package.  ``Compressible'' indicate  gas dynamics, ISPH: Incompressible SPH, WCSPH: Weakly-Compressible SPH, Elastic: Elastic dynamics, DEM: Discrete element method, ``Astrophysics'' indicates that the code can compute gravitational acceleration.  }\\
  \midrule
  \textbf{AQUAGPUsph}~\cite{cercos2015aquagpusph}         & MPI \& GPU   & C++       & Cell-grid                       & No & No  & WCSPH\\
  \textbf{DualSPHysics}~\cite{crespo2015dualsphysics}     & GPU          & C++       & Cell-grid                            & No  & Yes & WCSPH, Elastic, DEM\\
  \textbf{GADGET4}~\cite{springel2020simulating}          & MPI          & C++       & Octree                        & No  & Yes & Compressible, Astrophysics\\
  \textbf{GIZMO}~\cite{hopkins2015new}                    & MPI          & C         & Octree                       & No & Yes & Compressible, Astrophysics \\
  \textbf{GPUSPH}~\cite{bilotta2016GPUsph}                & MPI \& GPU     & C++       & Cell-grid                       & No  & Yes & WCSPH  \\
  \textbf{PHANTOM}~\cite{phantom:2018}                & MPI          & FORTRAN   & Octree                        & Yes & Yes & Compressible, Astrophysics \\
  \textbf{PySPH}                                          & MPI \& GPU   & Python    & Cell-grid, Octree                & Yes  & Yes & WCSPH, ISPH, Compressible, Elastic, DEM\\
  \textbf{SPHERA}~\cite{amicarelli2020sphera}             & No           & FORTRAN   & Cell-grid                       & No  & No & WCSPH \\
  \textbf{SPHERAL++}~\cite{soft:spheral++}             & MPI           & C++/Python   & Octree                       & No  & No & WCSPH, Compressible, Astrophysics, Elastic \\
  \textbf{SPHinXsys}~\cite{zhang2020sphinxsys}            & No           & C++       & Cell-grid                       & No  &  Yes & WCSPH, Elastic\\
  \textbf{SPlisHSPlasH}~\cite{koschier_2019}              & MPI \& GPU   & C++/Python       & Cell-grid                  & No  & No & WCSPH, ISPH, Elastic \\
  \bottomrule
\end{tabular}
\caption{The summary of packages available for particle methods.}\label{table:compare}
\end{table}





From the Table~\ref{table:compare} it can be seen that all of these packages
(except PySPH) are implemented in a lower-level programming language like C,
C++, or FORTRAN. While some of these feature a Python interface, the primary
programming language used to implement the schemes is still in a lower-level
language. One can also see that apart from the PHANTOM package, the other
packages do not employ any form of continuous integration testing that is
public. Studies~\cite{prechelt_empirical_2000,gmys_comparative_2020} have
indicated that higher-level programming languages like Python~\cite{py:python}
lead to more programmer productivity. Python is also acknowledged by these
studies and other articles~\cite{py:nature:2015} as a popular language and as
a good introductory programming language~\cite{py:teaching-us}. Our own
experience has indicated that the use of a high-level language makes the code
a lot more accessible to students and researchers. Python features extensive
libraries for scientific computation, general purpose tasks, and easily
interfaces with lower-level languages like C/C++, FORTRAN as well as OpenCL and
CUDA. These reasons make it an ideal choice for use in PySPH.

Many of the existing SPH related packages are developed with a few particular
SPH schemes or application domains in mind. For example, many of the packages
mentioned above do not solve elastic dynamics and while some solve
incompressible fluid mechanics problems they do not support compressible fluid
flow and vice-versa. In addition, there are other meshless methods like the
Discrete Element Method (DEM)~\cite{cundall1979discrete}, Reproducing Kernel
Particle Method (RKPM)~\cite{liu:rkpm:95}, Moving Least-Squares Particle
Hydrodynamics (MLSPH)~\cite{dilts:mls:99}, and others
(see~\cite{nguyen:review:2008} for a review of meshless methods). Researchers
are continuing to explore new methods and techniques in order to improve their
efficiency and accuracy. It is difficult to perform research in a new area
without investing a significant amount of time in building the necessary
tools. Our goal in creating PySPH is to make it possible for researchers to
build upon the work of others with relative ease. To this end PySPH provides,
\begin{itemize}
\item data-structures to store particle properties irrespective of the
  meshless method being used.
\item algorithms to find nearest neighbors of particles for various uses.
\item the ability to write high-level Python code to express the
  inter-particle interactions and have them seamlessly execute on multi-core
  CPUs and GPUs.
\item implementation in the form of reusable libraries. This is in contrast
  with tools that only seek to solve specific problems.
\end{itemize}
Furthermore, to promote usability, extensibility, and quality, PySPH
\begin{itemize}
\item is distributed under a liberal, open-source, BSD (three-clause) license.
  This does not restrict commercial use of the software.
\item is a cross platform package that works on the major platforms today
  viz.\ Linux, Windows, and MacOS.
\item is written using modern software engineering practices including public
  hosting on \url{https://github.com/pypr/pysph}, easily accessible
  documentation, the use of unit tests, and continuous integration testing.
\item provides a large number of well established schemes.
\item provides an extensive collection of reusable benchmark and example
  problems.
\item uses a high-level and easy to read programming language,
  Python~\cite{py:python}.
\end{itemize}

In order to promote reproducible research, PySPH is relatively easy to install
using standard tools available in the Python ecosystem. This allows
researchers to share their relevant changes alone which in combination with
PySPH can be used to reproduce results. PySPH seeks to offer these features
without sacrificing computational performance. This is important since SPH
simulations can be computationally demanding. PySPH therefore runs on
multi-core CPUs, GPUs and also supports distributed execution on clusters.
PySPH supports these without requiring any major changes to the user code.
PySPH was originally designed to run in serial on CPUs and be distributed
using MPI. In 2013 it was redesigned and reimplemented to make use of
automatic code generation from high-level Python. This made PySPH much easier
to use and also demonstrated improved performance. Some historical details and
design decisions are mentioned in~\cite{PR:pysph:scipy16}.

PySPH uses object-orientation for maximum reuse and managing complexity.
Python is used to specify the inter-particle interactions and high-performance
code is generated from this high-level specification. Source templates are
used to abstract any common patterns and generate code. These features allow
for a great deal of flexibility without sacrificing much performance.

While PySPH has been used extensively by our group, it has been used in
several recent papers by other researchers in the
community~\cite{meyer_parameter_2020,arai_comparison_2020,li_smoothed_2018,bao_modified_2019}.

The next section outlines the basic equations employed in SPH in order to
provide a brief background. We then discuss the design of the PySPH framework
by showing how one may solve a simple problem with it. We use this example to
highlight the features and design of PySPH. We then show several sample
applications from PySPH to benchmark and demonstrate its wide applicability.
We also discuss the computational performance of PySPH.


\section{The SPH method}\label{sec:sph}

In this section we briefly introduce the SPH method. We refer the reader
to~\cite{violeau_book_2012} for a detailed discussion. SPH is a Lagrangian,
meshless, particle method, where the particles represent the underlying
continuum. In SPH, a field variable $f(\ten{x})$ over a domain
$\Omega$ can be approximated by a convolution with the kernel function
$W(\ten{x}, h)$:
\begin{equation}
  \label{eq:sph:interpolation}
  f(\ten{x}) = \int_{\Omega} f(\ten{y}) W(\ten{x} - \ten{y}, h) \text{d}\ten{y},
\end{equation}
where $h$ is the smoothing length, and $\text{d}\ten{y}$ is the differential
volume element. The kernel function approximates the delta function in the
limit $h \rightarrow 0$ and is a compact function such that $W(\ten{r}) = 0$
when $|\ten{r}| > kh$ for some constant $k$. To obtain the discrete counterpart the
field is discretized into particles having mass $m_i$, density $\rho_i$, and
the differential volume is replaced by the volume $V_i = m_i/\rho_i$ where the
subscript denotes the index of the $i^{th}$ particle at position $\ten{x}_i$.
The integral is then approximated by a summation:
\begin{equation}
  \label{eq:sph:interpolation:sum}
  f(\ten{x}_i) = \sum_{j} V_j f(\ten{x}_j) W(\ten{x}_i - \ten{x}_j, h),
\end{equation}
where the summation is over all the particle in the support of the kernel
function.  If the function $f(\ten{x}_i)$ is replaced by density
$\rho(\ten{x}_i)$ we get the summation density formula in SPH,
\begin{equation}
  \label{eq:2}
  \rho(\ten{x}_i) = \sum_{j} m_j W(\ten{x}_i - \ten{x}_j, h).
\end{equation}
The SPH approximation can be used to discretize partial differential equations.
The governing equations of motion for a fluid flow, expressed in the Lagrangian
form, are,
\begin{equation}
  \label{eq:ns:mass}
   \frac{d\rho}{dt} = - \rho \nabla \cdot \ten{u},
\end{equation}
\begin{equation}
  \label{eq:ns:mom}
  \frac{d\ten{u}}{dt} = - \frac{1}{\rho}\nabla p +
  \nu \nabla^2 \ten{u} + \ten{g},
\end{equation}
where $\frac{d(.)}{dt}$ is the material derivative, $\rho$ is the density, $p$
is the pressure, $\nu$ is the kinematic viscosity, and $\ten{g}$ is the
external acceleration. In the case of a weakly-compressible fluid, the
equation of state which completes the above equations is,
\begin{equation}
  \label{eq:eos}
  p = \frac{c^2_0 \rho_0}{\gamma}
  \left[{\left(\frac{\rho}{\rho_0}\right)}^{\gamma} - 1\right],
\end{equation}
where $\gamma = 7$, $\rho_0$ and $c_0$ is the reference density and speed of
sound. In the case of a compressible fluid, the equations include an energy
equation and a different equation of state.  The discrete form of the above
equations using SPH approximation can be written as,
\begin{equation}
  \label{eqn:continuity:sph}
  \frac{d\rho_i}{dt} = \sum_{j} m_j (\ten{v}_i - \ten{v}_j) \cdot
  \nabla_i W_{ij},
\end{equation}
\begin{equation}
  \label{eq:mom:sph}
  \frac{d\ten{u}_i}{dt} = \ten{a}_{\text{p}} + \ten{a}_{\text{visc}}
  + \ten{a}_{\text{body}},
\end{equation}
where $\rho_i$, $\ten{u}_i$ is respectively the density and velocity of the
$i^{\text{th}}$ particle. $\ten{a}_{\text{p}}$ is the acceleration due to
pressure gradient commonly discretized~\cite{monaghan92} as,
\begin{equation}
  \label{eq:mom:sph:press}
  \ten{a}_{\text{p}} = - \sum_{j} m_j
  \left( \frac{p_i}{\rho_i^2} + \frac{p_j}{\rho_j^2} \right) \nabla_i W_{ij},
\end{equation}
where $p_i$ is pressure of $i^{\text{th}}$ particle, $p_j$ is pressure, $m_j$ is
the mass, and $\rho_j$ is the density of the $j^{\text{th}}$ particle,
$\nabla_i W_{ij} = \frac{\partial W}{\partial \ten{x}_i}(\ten{x}_i - \ten{x}_j,
h)$ is the gradient of the smoothing kernel. $\ten{a}_{\text{visc}}$ is the
acceleration due to viscous forces is~\cite{monaghan-review:2005},
\begin{equation}
  \label{eq:mom:sph:visc}
  \ten{a}_{\text{visc}} = \sum_{j} m_j \frac{4 \nu}{(\rho_i + \rho_j)}
  \frac{\ten{r}_{ij} \cdot \nabla_i W_{ij}}{(|\ten{r}_{ij}|^{2} + \eta
      h^{2})} \ten{u}_{ij},
\end{equation}
where $\ten{r}_{ij} = \ten{x}_i - \ten{x}_j$, and $\ten{a}_{\text{body}}$ is the
acceleration due to external body forces. Finally, the particle positions are
obtained by integrating in time the following ODE,
\begin{equation}
  \label{eq:eom}
  \frac{d \ten{x}_i}{dt} = \ten{u}_i.
\end{equation}
The problem we are going to consider in the next section is the simulation of
a two-dimensional dam break~\cite{wcsph-state-of-the-art-2010} where a column
of fluid is placed in a tank and simulated under the influence of gravity. For
the purpose of illustration we consider a simple treatment of the boundary
conditions. The tank is represented by solid particles having the same mass as
the fluid particles on which we solve the continuity equation and then set the
pressure by using the equation of state equation~\eqref{eq:eos}.

\subsection{Time Integration}
The equations of motion and momentum equations are then integrated in time to
obtain the positions and velocities respectively, a Predictor-Corrector
integrator~\cite{monaghan-review:2005} is used to integrate in time which is
described as,
\begin{equation}
  \ten{u}^{n+\frac{1}{2}}_i = \ten{u}^{n}_i + \frac{\Delta t}{2}{\left(\frac{d
              \ten{u}_i}{dt}\right)}^n,
\end{equation}
\begin{equation}
  \ten{x}^{n+1}_i = \ten{x}^{n}_i + \Delta t\ten{u}^{n + \frac{1}{2}}_i,
\end{equation}
\begin{equation}
  \ten{u}^{n+1}_i = \ten{u}^{n+\frac{1}{2}}_i + \frac{\Delta
    t}{2}{\left(\frac{d \ten{u}_i}{dt}\right)}^{n + 1}.
\end{equation}
The time-step for the integration is based on a CFL criterion that depends on
the differential equations being solved.  In the SPH literature, implicit
time-stepping is not very common as a result PySPH does not currently support
these. The issue of stiffness of the system is also not explicitly handled by
PySPH. However, it is useful to note that PySPH does support iterative
solution of matrices using a matrix-free formulation and this is discussed in
a little more detail in section~\ref{sec:iterated_group}.

\subsection{Nearest neighbor particle search}
\label{sec:nnps}

\begin{figure}[!h]
  \centering
  \includegraphics[width=0.4\textwidth]{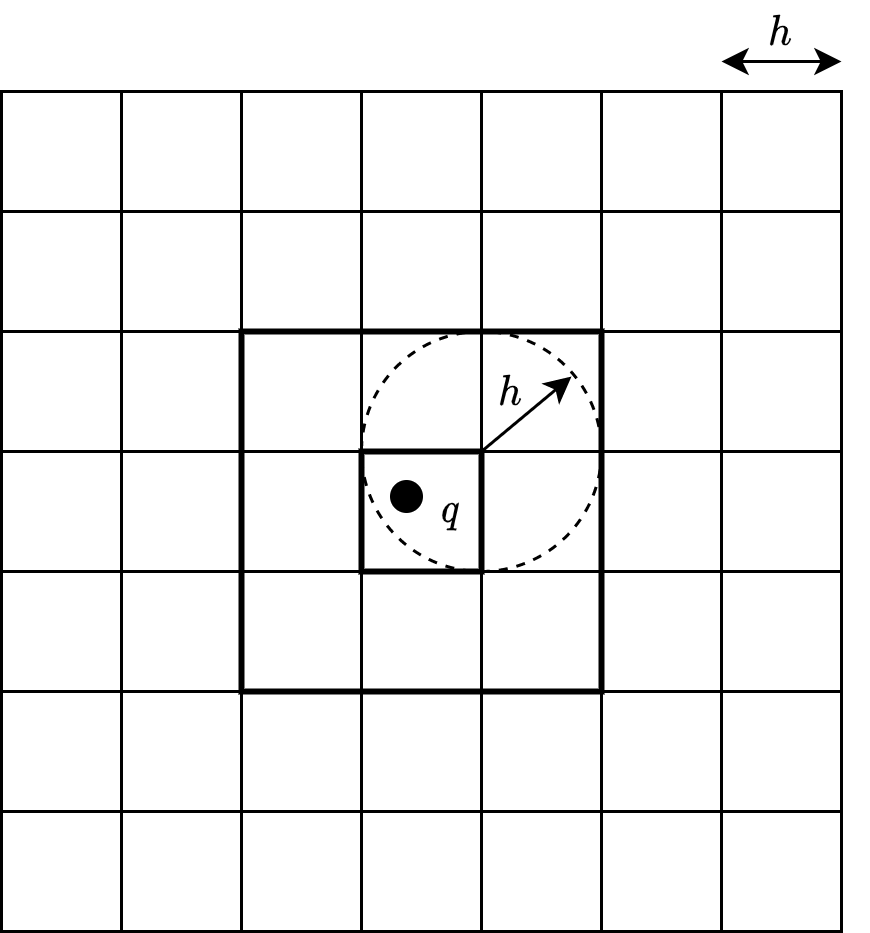}
  \caption{Nearest neighbors of particle $q$ with support radius $h$ can be
    found in the 9 neighboring boxes of the cell that $q$ belongs to.}
\label{fig:nnps}
\end{figure}

In order to find the particles that fall under the support of a kernel, a
nearest neighbor particle searching (NNPS) algorithm is used. The uniform grid
data structure is the most commonly used approach for NNPS. It works by
dividing the particles into cells of size equal to the support radius. The
neighbors of a particle are then searched in the neighboring $3^d$ boxes of
the query particle as shown in figure~\ref{fig:nnps}, where $d$ is the number
of dimensions. This method works efficiently when the support radius is
constant for all particles, but when the support radius is variable, the cell
size needs to be the maximum support radius which makes the uniform grid based
approach inefficient as the cells become too large. To efficiently find
neighbors in cases of variable support radius, a data structure with adaptive
cell size needs to be used. The most common approach in these cases is to use
an Octree~\cite{Hernquist_1989,springel_2005}. As part of PySPH, we provide
the following algorithms for NNPS on the CPU:

\begin{sloppypar}
  \begin{itemize}
  \item \textbf{Spatial Hash}. In this method, each cell in the uniform
    grid is mapped to an index in a hash table. Using a hash table allows
    storing of only the non-empty cells at the cost of a poor cache
    performance.
  \item \textbf{Linked List}~\cite{Dominguez_2011}. This method maps each cell
    to a flattened 1D cell ID which is mapped into a \code{head} array which
    stores the first particle of that cell. Another array \codeblack{next}
    maps each particle to the next particle in that cell. This algorithm gives
    a better cache performance because of its compactness, but needs to
    allocate an array of size equal to the number of cells in the domain
    including the empty cells.
  \item \textbf{Z-Order Space Filling Curve}. Space filling curves are used to
    map 3D and 2D data to one-dimension while preserving spatial locality. A
    Z-Order SFC is constructed by sorting the particle IDs by keys that are
    generated using bit-interleaving~\cite{morton_1966}. This method gives the
    best cache performance in uniform grids, but requires a sort in the build
    step which makes it slower overall than the linked list NNPS.
  \item \textbf{Octree}. In this method we recursively divide the space into 8
    octants until a node has less than a specified number of particles. When
    querying neighbors, the tree is traversed in a top-down order and only the
    octants that intersect with the bounding box of the query region are
    traversed.
  \end{itemize}
\end{sloppypar}
In addition to these, there are a few experimental NNPS algorithms implemented
in PySPH which will be discussed in a later publication. Due to its highly
parallel nature and a number of other restrictions, not all of the above
algorithms can be efficiently implemented on a GPU. PySPH provides the Z-Order
SFC, and the Octree algorithms for NNPS on the GPU.

\subsection{Schemes}

The SPH method can be used to simulate a wide variety of problems. There are
many schemes and variants of the basic scheme. The PySPH framework provides an
implementation of many of these.
\begin{itemize}
\item In the weakly compressible SPH family, PySPH implements the classic
  scheme of \cite{sph:fsf:monaghan-jcp94}, the corrections
  from~\cite{hughes-graham:compare-wcsph:jhr:2010}, the tensile instability
  correction~\cite{sph:tensile-instab:monaghan:jcp2000}, the
  $\delta$-SPH~\cite{marrone-deltasph:cmame:2011} method, the Transport
  Velocity Formulation (TVF)~\cite{Adami2013}, the Generalized TVF
  (GTVF)~\cite{zhang_hu_adams17}, the Entropically Damped Artificial
  compressibility (EDAC)~\cite{edac-sph:cf:2019}, the Dual time SPH
  (DTSPH)~\cite{pr:dtsph:2019}. Equations modeling surface
  tension~\cite{morris:surface:tension:2000, adami:surface:tension:2010} are
  also included in the framework.

\item In the family of Incompressible SPH (ISPH) schemes, the projection
  method of~\cite{sph:psph:cummins-rudman:jcp:1999}, the Implicit ISPH
  (IISPH)~\cite{iisph:ihmsen:tvcg-2014}, the simple iterative ISPH
  (SISPH)~\cite{muta2019simple}.

\item In the family of compressible SPH schemes, the Godunov SPH
  (GSPH)~\cite{inutsuka2002}, and the approximate Riemann solver with Godunov
  SPH~\cite{puri2014}, the adaptive density kernel estimate (ADKE)
  method~\cite{sigalotti2006}, the Monaghan-Price-Morris (MPM)
  method~\cite{price_2012}, and the conservative reproducible kernel
  SPH~\cite{crksph:jcp:2017}.

\item For elastic dynamics PySPH implements the method of
  Gray-Monaghan~\cite{gray2001}, the Generalized TVF
  (GTVF)~\cite{zhang_hu_adams17}, two way rigid-fluids
  coupling~\cite{akinci2012} and rigid body dynamics.

\item For the shallow water equations, PySPH implements the dynamic particle
  splitting and merging~\cite{vacondio_shallow_2013, vacondio_accurate_2012},
  Corrected SPH (CSPH)~\cite{rodriguezpaz_2005}, and the Lax-Friedrichs flux based
  stabilization term~\cite{ata_2005}.

\item Additional improvements and corrections that are available are the
  kernel and gradient corrections of~\cite{bonet_lok:cmame:1999} and the
  shifting algorithms
  of~\cite{acc_stab_xu:jcp:2009,fickian_smoothing_sph:skillen:cmame:2013}.

\item Boundary conditions are also implemented in the framework which
  include, wall boundary conditions~\cite{Adami2012}, inlet boundary
  conditions and various outlet boundary conditions which includes, Do
  nothing~\cite{FEDERICO201235}, Hybrid~\cite{negi2019improved}, Method of
  characteristics~\cite{Lastiwka2009:nonrefbc}, Mirror~\cite{tafuni2018}
  boundary conditions.

\end{itemize}

We next discuss the implementation, usage and design of PySPH in considerable
detail.


\section{Implementation, usage, and design}
\label{sec:design}

PySPH may be installed using standard installation procedures available in the
Python ecosystem. Documentation on this is available as part of the PySPH
online documentation. The documentation is available at
\url{https://pysph.readthedocs.io} and includes installation, tutorial, and
reference material on PySPH.

\subsection{Design overview}
\label{sec:design-overview}

The overall design of PySPH can be broken down into the following pieces of
infrastructure.

\begin{description}
\item[\texttt{pysph.base}] provides utilities for the management of a large
  number of particles.

  \begin{itemize}
  \item The package provides a \code{ParticleArray} class which contains the
    functionality to create and remove particles, access and change their
    properties etc.
  \item The \code{pysph.base} package also implements many algorithms to find
    the nearest neighbors of particles in the vicinity of a given particle.
    These algorithms include ones for fixed and variable smoothing kernel
    radius as well as CPU and GPU algorithms as discussed in
    Section~\ref{sec:nnps}.
  \item The commonly used SPH kernels are also available in
    \code{pysph.base.kernels}.
  \end{itemize}

\item[\texttt{pysph.sph}] provides functionality to write the low-level
  computations for SPH in pure Python. Internal functionality to convert the
  high-level Python code describing the SPH algorithms into high-performance
  code is provided here. This includes execution on multi-core CPUs and GPU
  architectures. The building blocks of many existing SPH formulations are
  also implemented in this package as \code{Equation}s and \code{Scheme}s
  which are discussed later.
\item[\texttt{pysph.solver}] provides the \code{Application} and \code{Solver}
  classes that combine the basic tools in the other packages to create
  easy-to-use command line applications that solve specific problems.
\item[\texttt{pysph.tools}] provides many general purpose miscellaneous tools
  for visualization, post-processing, input/output, etc.
\item[\texttt{pysph.parallel}] contains the code for distributed execution via
  MPI.
\item[\texttt{pysph.examples}] provides over 90 examples that simulate many
  benchmark problems. Some of these include post-processing and comparison
  with exact, experimental, and computational results. These examples are
  written in a way that they may be reused.
\end{description}

In order to perform the above efficiently, PySPH relies on general purpose
functionality provided by the following sister packages.
\begin{itemize}
\item \code{cyarray} is a simple package that provides an interface to a typed
  C-like array.  It also interfaces easily with NumPy~\cite{numpy-guide} arrays.
\item \code{compyle} allows a user to generate high-performance code from a
  restricted subset of pure Python. This can be used to write algorithms that
  use multi-core CPUs and GPUs efficiently but from the convenience of the
  Python programming language. More details are available in
  \cite{compyle-scipy-2020}.
\item \code{pyzoltan} is a Python wrapper for the
  ZOLTAN\cite{zoltan-web,zoltan:2012} package. It is used to distribute the
  particles across multiple computers to scale up the
  computations~\cite{kunal-parcomptech-2013}.
\end{itemize}
These are currently separate packages but were originally part of the PySPH
package. They are general purpose packages and may be used independently of
PySPH.

\subsection{An example problem}
\label{sec:design-example}

PySPH provides many low-level tools for a developer to create their own
particle methods simulator. However, PySPH provides a high-level framework for
the convenience of researchers whose focus is on the rapid implementation of
an SPH code. In this section, this high-level framework is described. We look
at a simple example problem and show how to implement this in PySPH. We then
go over the details of how PySPH executes the resulting example internally.

We consider a simple dam-break problem in two dimensions with the standard
weakly-compressible SPH formulation~\cite{sph:fsf:monaghan-jcp94} whose
equations are discussed in the earlier section. This example is available as
part of the repository\footnote{Repository for the paper is at
  \url{https://gitlab.com/pypr/pysph_paper}.} supporting this article,
\href{https://gitlab.com/pypr/pysph_paper/-/blob/master/code/db2d.py}{here}.
The file is about 125 lines of pure Python. We can immediately execute the
example using the following:
\begin{verbatim}
$ python db2d.py
\end{verbatim}
This will generate high-performance C++ code that runs on a single core of a
CPU. As the simulation runs, output files are dumped into a \code{db2d_output}
directory. The output files are in the form of NumPy NpzFiles or HDF5 files
(if the package \code{h5py} is installed).

We may run the same code on all CPU cores with this invocation,
\begin{verbatim}
$ python db2d.py --openmp
\end{verbatim}
Furthermore, if one has a GPU or OpenCL support, one can also invoke this as,
\begin{verbatim}
$ python db2d.py --opencl
\end{verbatim}

If one is using MPI and has configured PySPH to use MPI, one may run this as,
\begin{verbatim}
$ mpirun -np 4 python db2d.py
\end{verbatim}
No change to the code is needed. Although with the very few particles in this
problem, it would be much slower to execute this on a GPU or on multiple CPUs
with MPI. The ability for PySPH to execute the same unmodified Python code on
different platforms efficiently is an important and powerful feature.

Thus, the Python file is really a specification for the SPH simulation and the
framework takes care of generating the appropriate code. Since the code is
Python, it promotes reuse and one may use the standard Python mechanisms for
this.

The default application also supports many command line options, these may be
seen by passing the \code{--help} option to the program which will print a
help message with all command line options.

We now look at the details of how we implement this. The high-level PySPH
framework employs an object-oriented design to minimize repetition and
facilitate reuse. The first step in solving the dam-break problem is to create
a new \code{Application} subclass that implements the necessary methods. We
show an outline of such an example in the following
listing~\ref{lst:example1}.

\begin{lstlisting}[label={lst:example1},caption={Outline of 2d dam break example.}]
from pysph.solver.application import Application

class DamBreak2D(Application):
    def create_particles(self):
        # ...
    def create_equations(self):
        # ...
    def create_solver(self):
        # ...

if __name__ == '__main__':
    app = DamBreak2D()
    app.run()
\end{lstlisting}

The code in Listing~\ref{lst:example1} outlines the basic methods that one may
overload to create an application. The important methods shown are the
following:
\begin{itemize}
\item \code{create_particles}: where the user creates \code{ParticleArray}
  instances to represent the different physical entities as particles.

\item \code{create_equations}: where the SPH equations are suitably grouped
  and instantiated. These equations typically implement the right-hand-side of
  the SPH discretized equations~\eqref{eqn:continuity:sph},
  \eqref{eq:mom:sph}.

\item \code{create_solver}: where a \code{Solver} instance is suitably
  created. The solver manages an \code{Integrator} which in turn uses the
  generated equations to compute the acceleration terms.
\end{itemize}

We now show the details of the various methods. We begin with the creation of
the particles.

\subsubsection{Creation of particles}
\label{sec:create_particles}

\begin{lstlisting}[label={lst:ex:create_particles}, caption={Implementation of
the \code{create_particles} method.}]
import pysph.tools.geometry as G
from pysph.base.utils import get_particle_array_wcsph

class DamBreak2D(Application):
    def create_particles(self):
        dx, hdx, rho = 0.1, 1.2, 1000
        xf, yf = G.get_2d_block(
            dx=dx, length=1.0, height=2.0, center=[-1.5+dx, 1+dx]
        )
        m = dx**2 * rho
        fluid = get_particle_array_wcsph(
            name='fluid', x=xf, y=yf, h=hdx*dx, m=m, rho=rho
        )
        xt, yt = G.get_2d_tank(
            dx=dx, length=4.0, height=4.0, num_layers=3
        )
        solid = get_particle_array_wcsph(
            name='solid', x=xt, y=yt, h=hdx*dx, m=m, rho=rho
        )
        return [fluid, solid]
\end{lstlisting}

In the Listing~\ref{lst:ex:create_particles}, the code used to generate the
geometry is shown. Here we make use of some functions provided in the
\code{pysph.tools.geometry} module to create a tank and a block of particles.
These functions return NumPy~\cite{numpy-guide} arrays that define the
particle positions. The \code{fluid} variable is a \code{ParticleArray}
instance. The \code{get_particle_array_wcsph} method is merely a convenience
that pre-defines several important default particle properties like the
\code{x, y, z, u, v, w, p} which represent the position, velocity, and
pressure. The property \code{h} is the smoothing length of the kernel and this
parameter also determines the search radius for the NNPS when finding
neighbors. It is important to note that all particle arrays should have a name
that is a valid Python variable name. The \code{solid} particle array is also
created similarly. The \code{pysph.tools.geometry} provides a few functions to
create shapes used in standard benchmark problems in two and three dimensions.
The \code{geometry} module has just been shown here for illustration and one
may create the particles in any other way as a collection of points specified
as NumPy arrays. Once the particle arrays are setup, they are returned from
the \code{create_particles} method as a list. A \code{ParticleArray} manages a
collection of properties and these are accessible in the manner shown in the
following example:

\begin{lstlisting}
>>> fluid.get_number_of_particles()
231
>>> fluid.x
array([-1.9, ..., -0.9])
>>> fluid.rho
array([1000., ..., 1000.])
\end{lstlisting}

Particle arrays allow us to operate on the entire collection of particles as a
whole. This means that a property is something that has a potentially
different value for each particle. One may add more properties using the
\code{fluid.add_property()} method to the existing particle arrays. Each
property illustrated in the code (\code{x, y, u, v}) has a single quantity per
particle. For a vectorial or tensor quantities it is sometimes more convenient
to associate 3 or 9 elements per particle as a single property. For these
cases, particle arrays allow the specification of a ``stride''. Hence for a
vector quantity the stride would be set to 3. The data is still stored as a
one-dimensional array but the user could use the stride when accessing the
appropriate values.

Some properties like viscosity are constant for all the particles and can be
added using \code{fluid.add_constant()}. In addition to this there are methods
to add particles, remove particles, extract particles into another particle
array etc. The key idea with particle arrays is that they allow us to
represent distinct entities in the problem being solved and associate any
number of properties or constants with these. These properties and constants
may be operated upon by the equations and integrators which are described
next.

We note that the \code{pysph.tools.read_mesh} module provides some simple
functionality to read input geometry data in the form of an input file that
describes a triangulation of a surface. The \code{mesh2points} function does
this given a suitable desired spacing of points. This is done through the use
of the \code{meshio} package\footnote{See
  \url{https://github.com/nschloe/meshio}} which supports many different file
formats. This functionality is currently quite basic but illustrates what is
possible.

\subsubsection{Setting up the particle interactions}
\label{sec:equations}

Now that the particle arrays are defined, we need to specify the
inter-particle interactions and this is done by the \code{create_equations()}
method of the application. Before we discuss this in detail, we provide a
broad overview.

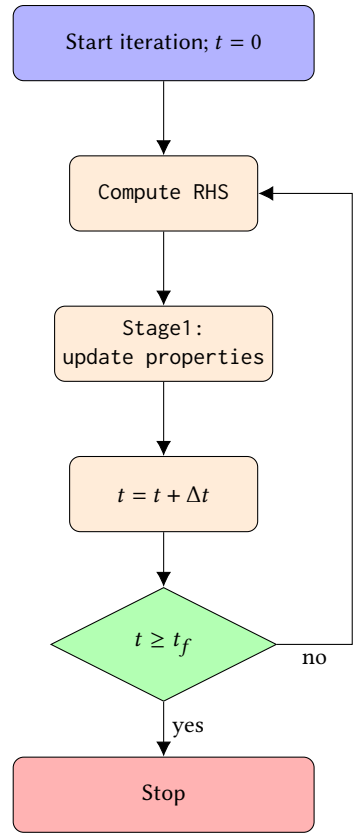
\begin{figure}[h!]
  \centering
  \begin{tikzpicture}[node distance=2cm,
    every node/.style={fill=none}, align=center
    ]
    \node (start) [activityStarts]{Start iteration; $t=0$};
    \node (rhs) [process, below of=start] {Compute RHS};
    \node (stage1) [process, below of=rhs] {Stage1:\\ update properties};
    \node (update_time) [process, below of=stage1] {$t = t + \Delta t$};
    \node (check) [decision, below of=update_time]
    {$t \geq t_f$};
    \node (stop) [startstop, below of=check] {Stop};

    \draw[->] (start) -- (rhs);
    \draw[->] (rhs) -- (stage1);
    \draw[->] (stage1) -- (update_time);
    \draw[->] (update_time) -- (check);
    \draw[->] (check) -- node[anchor=west] {yes} (stop);
    \draw[->] (check.east) -- node[anchor=north] {no} ++ (1,0) |- (rhs.east);
  \end{tikzpicture}

  \caption{Overview of a simple Euler integrator and the steps involved.}
  \label{fig:euler-integrator}
\end{figure}

In order to solve the dam-break problem, we need to first discretize the
entities involved and this has been performed in the previous section (see
section~\ref{sec:create_particles}). The equations we must solve are the
continuity equation~\eqref{eqn:continuity:sph}, which provides an ordinary
differential equation (ODE) for the density of each particle. The equation of
state, equation~\eqref{eq:eos} which relates the pressure to the density and
the momentum equation~\eqref{eq:mom:sph}, and \eqref{eq:mom:sph:press}.
Finally, we must move the particles according to equation~\eqref{eq:eom}. The
equations are all ODEs and we must first compute the right-hand-sides (RHS) of
these ODEs. In order to keep the example simple, we use a simple forward Euler
integrator. The steps involved in this are as shown in
Fig.~\ref{fig:euler-integrator}. The implementation of the integrator is
discussed in the Section~\ref{sec:integrator}. The inter-particle interactions
are all computed in the block titled ``Compute RHS''.

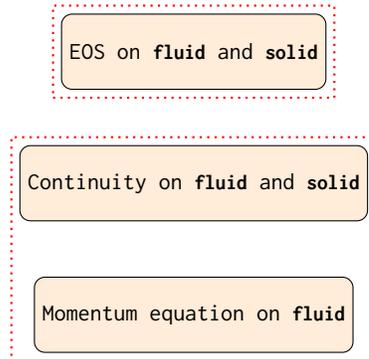
\begin{figure}[h!]
  \centering
  \begin{tikzpicture}[node distance=1.75cm,
    every node/.style={fill=white}, align=center]
    \node (eos) [process] {EOS on \code{fluid} and \code{solid}};
    \node (ce) [process, below of=eos] {Continuity on \code{fluid} and \code{solid}};
    \node (me) [process, below of = ce] {Momentum equation on \code{fluid}};

    \node [fit=(eos),draw,dotted,red, thick,fill=none] {};
    \node [fit=(ce) (me),draw,dotted,red,thick, fill=none] {};
  \end{tikzpicture}

  \caption{The inter-particle evaluations. The equation of state (EOS) is
    computed first for all particles before the continuity and momentum are
    evaluated in order to compute the correct pressure for all particles. This
    is denoted by the red dotted line grouping the individual steps.}
  \label{fig:equations}
\end{figure}

The steps involved in computing the RHS for this particular problem are as
shown in the Fig.~\ref{fig:equations}. It is to be noted that the momentum
equation requires the updated pressure of all the particles. The pressure is
computed using the equation of state (EOS), equation~\eqref{eq:eos}. We
therefore must complete the evaluation of the EOS on all particles before
computing the momentum equation. This is shown in the figure with a red dotted
line around each group of computations. Note that the RHS for both the
continuity and momentum equations have no inter-dependencies and may be
computed together.

Given this background, we now discuss the implementation of the above in
PySPH. The Listing~\ref{lst:ex:create_equations}, shows the implementation of
the \code{create_equations} method which describes the inter-particle
interactions.

\begin{lstlisting}[label={lst:ex:create_equations}, caption={Implementation of
the \code{create_equations} method.}]
# ...
from pysph.sph.equation import Equation, Group

class DamBreak2D(Application):
    def create_particles(self):
        # ...

    def create_equations(self):
        equations = [
            Group(equations=[
                EOS(dest='fluid', sources=None, rho0=1000, c0=10),
                EOS(dest='solid', sources=None, rho0=1000, c0=10),
            ]),
            Group(equations=[
                ContinuityEquation(
                    dest='fluid', sources=['fluid', 'solid']
                ),
                ContinuityEquation(
                    dest='solid', sources=['fluid']
                ),
                MomentumEquation(
                    dest='fluid', sources=['fluid', 'solid']
                )
            ])
        ]
        return equations
\end{lstlisting}

In the above we have not defined the \code{EOS}, \code{ContinuityEquation} or
the \code{MomentumEquation}. These are discussed later. However, we can see
the structure with which the equations are defined and expressed. The
structure is similar to that seen in Fig~\ref{fig:equations}. Each equation
has one \code{dest} and a list of \code{sources} with some additional
arguments which are used by the equation in calculations. The \code{dest} is
the name of the destination particle array for which we want to calculate the
RHS. The \code{sources} contains the particle array names which influence the
destination according to the equation in SPH form. The \code{Group} class
merely ensures that its equations are completed first before the next group
is executed. This is important as we must compute the pressure on all
particles \emph{before} the momentum equation is executed as discussed
earlier.

The equations allow users to specify the inter-particle interactions expressed
using certain conventions.  We demonstrate the equations implemented here.

\begin{lstlisting}[label={lst:equations},caption={Implementing equations.}]
class EOS(Equation):
    def __init__(self, dest, sources, rho0, c0, gamma=7):
        self.rho0 = rho0
        self.gamma = gamma
        self.B = rho0*c0*c0/gamma
        super(EOS, self).__init__(dest, sources)

    def initialize(self, d_idx, d_p, d_rho):
        tmp = pow(d_rho[d_idx]/self.rho0, self.gamma)
        d_p[d_idx] = self.B*(tmp - 1.0)


class ContinuityEquation(Equation):
    def initialize(self, d_idx, d_arho):
        d_arho[d_idx] = 0.0

    def loop(self, d_idx, s_idx, d_arho, s_m, DWIJ, VIJ):
        vijdotdwij = (DWIJ[0]*VIJ[0] + DWIJ[1]*VIJ[1] + DWIJ[2]*VIJ[2])
        d_arho[d_idx] += s_m[s_idx]*vijdotdwij


class MomentumEquation(Equation):
    def initialize(self, d_idx, d_au, d_av, d_aw):
        d_au[d_idx] = 0.0
        d_av[d_idx] = -9.81
        d_aw[d_idx] = 0.0

    def loop(self, d_idx, s_idx, d_rho, s_rho, d_p, s_p,
             d_au, d_av, d_aw, s_m, DWIJ):
        tmp = (d_p[d_idx]/(d_rho[d_idx]**2) +
               s_p[s_idx]/(s_rho[s_idx]**2))
        d_au[d_idx] += -s_m[s_idx] * tmp * DWIJ[0]
        d_av[d_idx] += -s_m[s_idx] * tmp * DWIJ[1]
        d_aw[d_idx] += -s_m[s_idx] * tmp * DWIJ[2]

\end{lstlisting}

The Listing~\ref{lst:equations} shows the three different equations that are
instantiated in the \code{create_equations} method. The first is the
\code{EOS}. On inspection of the code, the equation takes two positional
parameters called \code{dest, sources} which are as explained before. The
other keyword arguments serve to set some constants that are used in the
\code{initialize} method. The \code{initialize} method is called internally by
PySPH for every particle of the corresponding \code{dest} particle array. For
example, for the case where the \code{'fluid'} is the destination, this is
called for each fluid particle and the index into this particle is passed by
convention as \code{d_idx}. All method arguments prefixed with a \code{d_} are
destination array properties. As can be seen in the two lines of the method,
the pressure of the particles are being set based on the density of the fluid.

The \code{ContinuityEquation} does not require any additional parameters. Its
\code{initialize} method simply sets the density ``acceleration'' (or the
right hand side of the continuity equation), \code{arho} property, to zero for
each particle. The \code{loop} method is called internally by PySPH for all
neighbors of each destination particle. Hence, in our case, all the fluid
neighbors for each particle will be identified and for each neighbor, the
method is called with \code{d_idx} being the destination index, and
\code{s_idx} being the source index. All arguments prefixed with \code{s_} are
source array properties. Thus \code{s_m} will be the fluid mass array when the
source is a fluid and the solid mass array when the solid is the source. All
the fluid neighbors are processed and then the solid neighbors are processed.
The special arguments to the \code{loop} method are \code{DWIJ} and
\code{VIJ}. These represent the kernel gradient and the difference between the
destination and source particle velocities as two 3-vectors. These occur
commonly in most SPH calculations that they are provided as a convenience when
the equation requests them in the arguments. They are only computed when
required by the user. There are several other standard pre-defined values that
are made available when required and are fully documented in the PySPH
documentation.

The only major change in the \code{MomentumEquation} is the use of the
\code{au, av, aw} properties that refer to the positional acceleration of the
particles. These equations completely describe the inter-particle interactions
and as can be seen are easy to write and do not require a large amount of
programming.

It is important to note that all of the equations shown above are not used in
their pure Python form. The high-level Python classes are internally compiled
into a suitable high-performance language and transparently executed. This
places some heavy restrictions on what can be written inside these methods.
One should not create any Python or NumPy objects inside these methods. The
idea is that the methods purely operate on properties and constants set in the
particle array. Any constants that are set as attributes of the class, for
example, \code{self.rho0, self.gamma} etc. may be used. All basic mathematical
operations are available but not any arbitrary Python function. The equations
therefore serve the purpose of specifying the inter-particle interactions
written out in terms of the particle properties and basic mathematical
operations. PySPH does allow users to write their own Python functions and use
them but we discuss these in greater detail in the PySPH documentation. An
outline of how the high-performance code is generated is discussed later in
Section~\ref{sec:detailed-design}.

\subsubsection{Setting up the integrator}
\label{sec:integrator}

The last part of the code is the application's \code{create_solver} method.
A simple implementation of this method is shown below.
\begin{lstlisting}[label={lst:ex:create_solver}, caption={Implementation of
the \code{create_solver} method.}]
from pysph.solver.solver import Solver
from pysph.base.kernels import CubicSpline

class DamBreak2D(Application):
    # other methods ...
    def create_solver(self):
        kernel = CubicSpline(dim=2)
        integrator = EulerIntegrator(
            fluid=EulerStep(), solid=EulerStep()
        )

        solver = Solver(
            kernel=kernel, dim=2, integrator=integrator,
            dt=2e-4, tf=1.0, pfreq=10
        )
        return solver
\end{lstlisting}

Listing~\ref{lst:ex:create_solver} demonstrates a simple \code{create_solver}
implementation. Here we instantiate the kernel to be used, PySPH provides many
standard kernels (which are also implemented in Python but are not shown
here). We import the standard \code{CubicSpline} kernel. We then create an
integrator which defines how the particles are to be integrated. We use a
\code{EulerIntegrator} and define a stepper which describes how the properties
are stepped in time. Finally, a \code{Solver} is created which puts these
together along with the time step, total time for integration, and returns the
instance. The \code{Solver} instance manages the time integration. Note that
here the time step is set directly whereas the time step is typically chosen
carefully based on stability considerations. The \code{pfreq} parameter is the
output frequency and in the above, output files are written every 10
iterations.

\begin{lstlisting}[caption={Implementation of integrators.},label={lst:integrator}]
class EulerIntegrator(Integrator):
    def one_timestep(self, t, dt):
        self.compute_accelerations()
        self.stage1()
        self.update_domain()
        self.do_post_stage(dt, 1)


class EulerStep(IntegratorStep):
    def stage1(self, d_idx, d_u, d_v, d_au, d_av,
               d_x, d_y, d_rho, d_arho, dt):
        d_u[d_idx] += dt*d_au[d_idx]
        d_v[d_idx] += dt*d_av[d_idx]
        d_x[d_idx] += dt*d_u[d_idx]
        d_y[d_idx] += dt*d_v[d_idx]
        d_rho[d_idx] += dt*d_arho[d_idx]
\end{lstlisting}

Listing~\ref{lst:integrator} shows the complete implementation of the
integrator and stepper. They use similar conventions with the properties all
prefixed with \code{d_} and the integrator stepper methods being called once
for each particle. The integrator provides a \code{compute_accelerations()}
method that computes the RHS using the equations provided. This method also
updates the neighbor information before computing the accelerations. The
\code{stage1()} method calls the stepper for each particle array defined, and
\code{update_domain()} performs any periodicity related modifications if
needed. The \code{do_post_stage} method allows users to attach callbacks to
perform operations after each stage is completed. Together, these implement
the interactions of the particles and the solution of the ODEs as shown in
Fig.~\ref{fig:euler-integrator}.

This is a complete example and the repository supporting this article
(\url{gitlab.com/pypr/pysph_paper}) includes a complete file that can be
executed. When executed, this solves the dam break problem with SPH. We have
only shown the features of PySPH necessary for this simple application. The
application features many other methods that may be used to configure the
application. For example, one may override \code{add_user_options} to add any
additional command line argument options, \code{create_domain} to set periodic
or mirror boundary conditions, \code{post_process} to post-process the
simulation data. The \code{pre_step, post_step, post_stage} methods are called
before or after each timestep or after each stage of the integrator is
complete respectively and execute any Python code. Finally, one may also
subclass the application we have just created, and override a few methods to
change its behavior.

As noted before, there are many SPH schemes available, for example WCSPH, and
EDAC-SPH. When the WCSPH scheme is used for different problems, one typically
needs to use the same set of equations in different combinations. PySPH
provides what are called \code{Scheme}s which abstract out the required code
to generate the equations and integrators. PySPH provides a variety of
standard schemes like the \code{WCSPHScheme}, \code{TVFScheme},
\code{EDACScheme}, \code{GSPHScheme}, etc.\ which allow users to solve a
problem by creating the particles and creating a suitably configured scheme. A
\code{SchemeChooser} is also provided that allows one to choose a suitable
scheme from the command line. These make it easy to create a single
application which can be run with a variety of different schemes. Many of the
PySPH examples use this approach.

\subsubsection{Visualization, and other features}
\label{sec:pysph-run}

The example application that was created in the previous section may be
immediately executed as discussed at the beginning of
Section~\ref{sec:design-example} for example,
\begin{verbatim}
$ python db2d.py --openmp
\end{verbatim}
will execute the example using OpenMP to utilize multiple cores on the
computer. As discussed earlier, the output is saved to the directory
\code{db2d_output}.

\begin{figure}[h!]
  \centering
  \includegraphics[width=\textwidth]{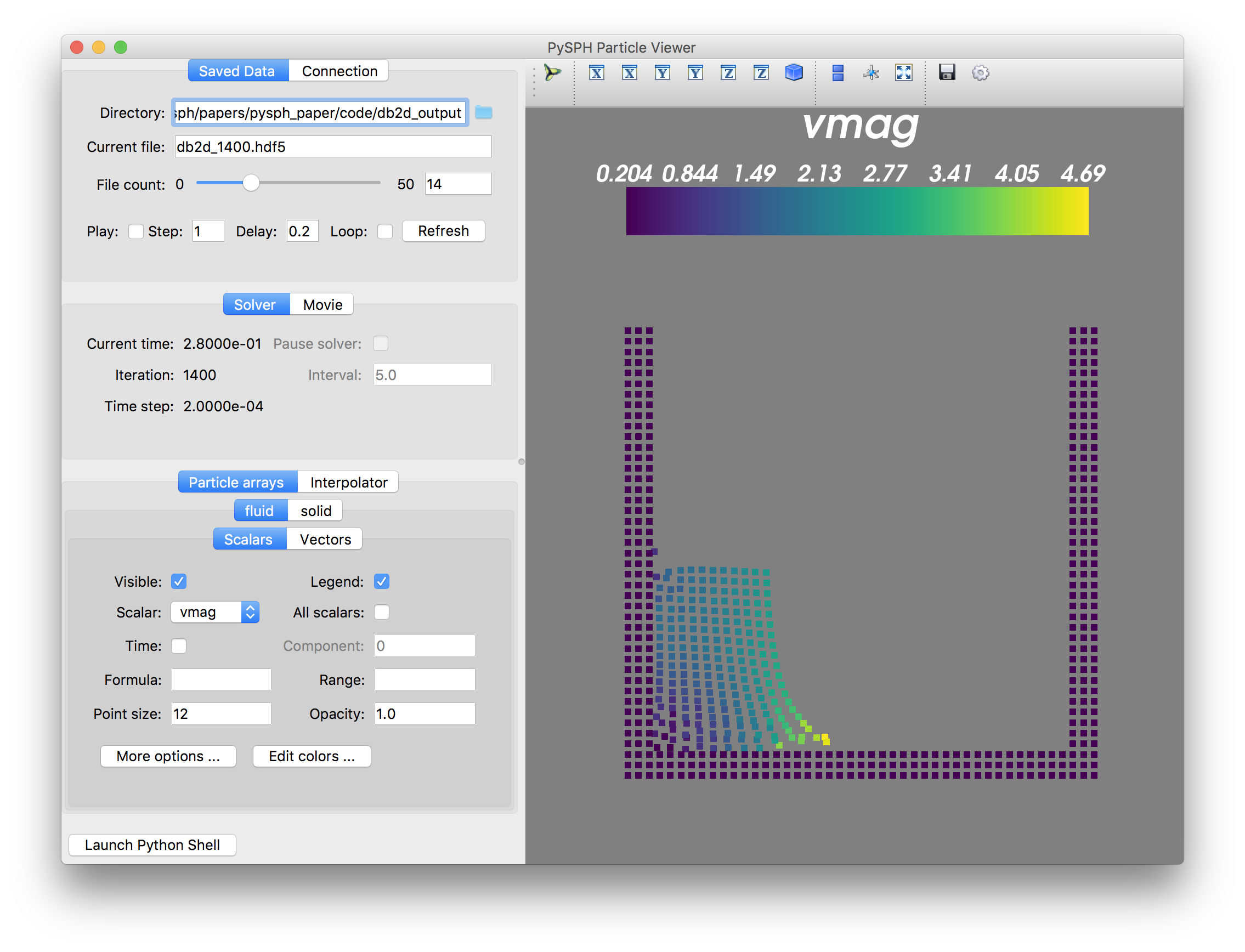}
  \caption{The built-in PySPH viewer visualizing a simple 2D dam-break
    problem. The color of the fluid particles indicates the velocity
    magnitude.}
  \label{fig:pysph-view}
\end{figure}

PySPH also ships with a simple 3D viewer that allows users to quickly
visualize the generated output. When PySPH is installed it installs a script
called \code{pysph}. The viewer is available as a sub-command of the
\code{pysph} command. The viewer can be used to visualize the results of the
simulation above using:
\begin{verbatim}
$ pysph view db2d_output
\end{verbatim}
The resulting viewer after some configuration is as shown in
Fig.~\ref{fig:pysph-view}. The viewer uses the
Mayavi~\cite{it:mayavi:cise:gael2011} Python package for 3D visualization and
presents a customized UI suitable for particle visualization. One may quickly
visualize a set of output files and view any number of particle arrays. The
files may be viewed as a timeseries and animated. Both scalars and vectors can
be viewed. In addition, the viewer may be customized with default
visualization settings by overriding the \code{Application.customize_output}.
This viewer is only suitable for desktops or workstations as it presents a
desktop UI. It provides an interactive environment suitable for up to around 5
million particles (for a typical modern laptop or desktop).

For massive parallel simulations with many more particles one may wish to use
other suitable visualization tools. The command \code{pysph dump_vtk} converts
the output data files into the VTK format.

\begin{figure}[h!]
  \centering
  \includegraphics[width=0.8\textwidth]{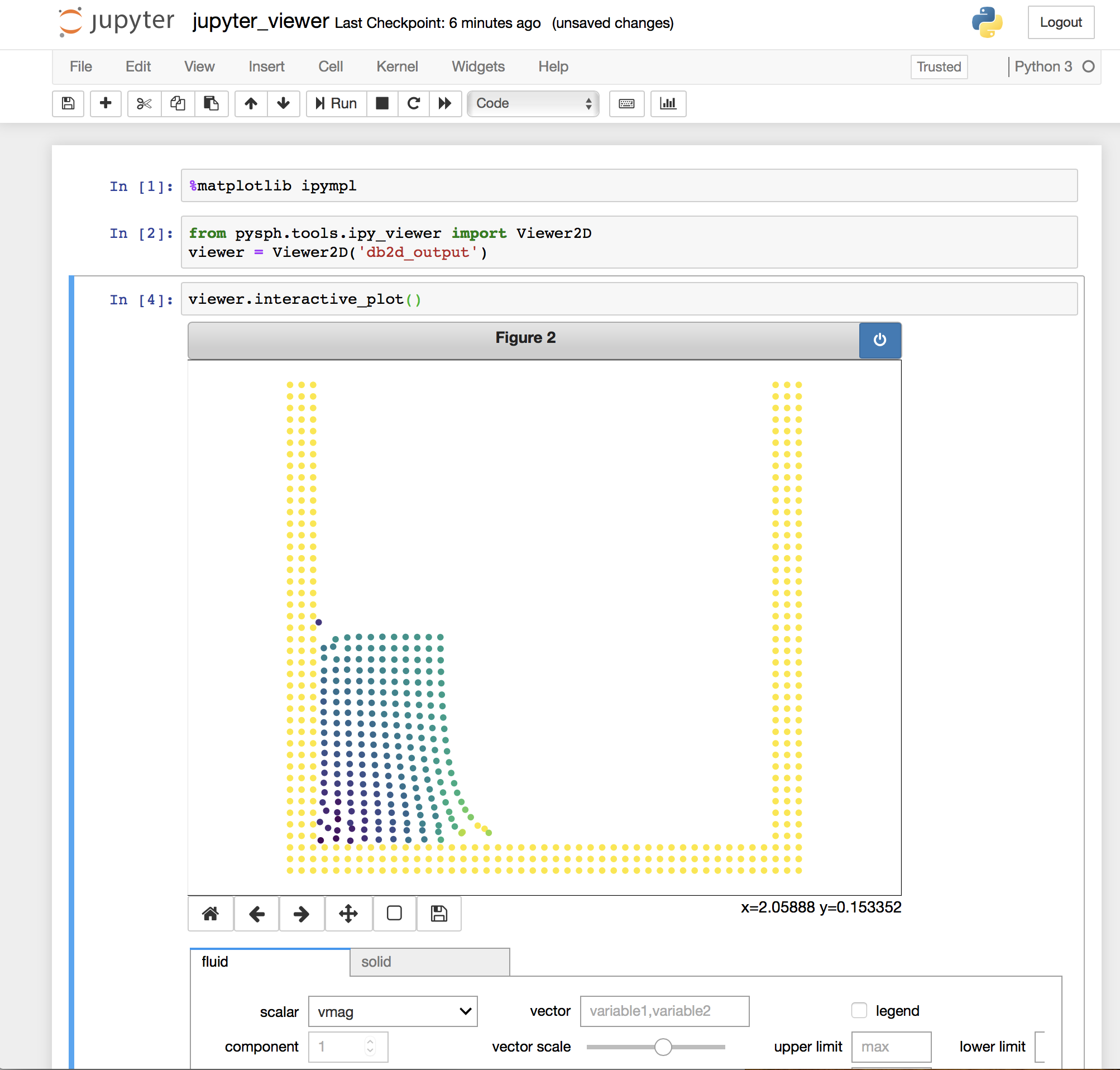}
  \caption{The jupyter notebook viewer showing the 2D dam-break problem
    results in a browser.}
  \label{fig:ipy-view}
\end{figure}

PySPH also features a viewer that allows one to share or visualize results
using a Jupyter notebook~\cite{jupyter:2016}. These can also be visualized on
the web using a browser. This is shown in the Fig.~\ref{fig:ipy-view}. It is
easy to share the results of a simulation online by hosting the files on a git
repository on \url{github.com} or \url{gitlab.com} using the Binder
project~\cite{binder:2018}. This is facilitated by the \code{pysph binder}
command which when given a directory, generates suitable Jupyter notebook
files as well as the other files required for the Binder project to work. One
can then share this with a colleague and have a full fledged Python
environment be created so anyone can view and interact with the data files.
They may also perform their own analysis on the data provided in these
environments.

We have created a repository with some sample results shown in this paper at
\url{https://gitlab.com/pypr/pysph_demo} which contains the simulation data
and can be viewed online using the binder launch link provided in the
\code{README}.

For more specialized analysis, the generated output can be easily loaded into
Python with a few lines of code.
\begin{lstlisting}
from pysph.solver.utils import load
data = load('db2d_output/db2d_4000.hdf5')
fluid = data['arrays']['fluid']
solid = data['arrays']['solid']
print(fluid.x.max())
\end{lstlisting}
Here, the \code{fluid} and \code{solid} variables are \code{ParticleArray}
instances with the saved values of the respective arrays. The last line simply
prints the maximum value of the $x$-coordinate of the fluid. Any kind of
post-processing and analysis may be performed using Python syntax.

In addition, PySPH provides many tools to perform a variety of post-processing
using the same infrastructure that is used to run the simulations. This can be
very powerful as shown in the following simple example,
\begin{lstlisting}
from pysph.tools.interpolator import Interpolator
interp = Interpolator([fluid, solid], num_points=10000)
p = interp.interpolate('p')
\end{lstlisting}
The above example interpolates the pressure field due to the two given
particle arrays, that were earlier read from the file, on a uniform mesh of
10000 points that encloses the domain. The interpolator uses the same
infrastructure as the PySPH simulations and may also be passed additional
parameters including a custom set of equations to change the nature of the
implementation or an arbitrary distribution of points to evaluate the values
on.

\begin{sloppypar}
  PySPH also provides a general purpose \code{SPHEvaluator} in the
  \code{pysph.tools.sph_evaluator} module. This may be passed a set of
  particle arrays and a collection of equations and a kernel. Internally,
  PySPH generates the high performance code, compiles it, and calls this code
  to evaluate the equations.
\end{sloppypar}
Thus, PySPH provides the basic building blocks to perform a variety of inter
particle computations very efficiently from the convenience of high-level
Python.

\subsection{Underlying design}
\label{sec:detailed-design}

Now that the basic features of PySPH have been understood from the perspective
of a user, we provide a high-level overview of how this is implemented. We
explain this in the context of the \code{Application.run()} method that is
called when one wishes to simulate any SPH problem (see
Listing~\ref{lst:example1}). The method internally performs the steps as
follows:
\begin{sloppypar}
  \begin{enumerate}
  \item Parsing of command line options passed to the script, setting up
    application logging, and passing the parsed command line arguments back to
    the user.

  \item Create the solver via \code{create_solver()}, thereafter the equations
    using \code{create_equations()}, and then the particles by calling
    \code{create_particles()}.

  \item Optionally create any periodic or mirror boundary condition domains,
    and any inlets and outlets.

  \item Instantiate a suitable nearest neighbor particle search (NNPS) algorithm
    to be used for the problem. PySPH features a variety of different NNPS
    algorithms as discussed in Section~\ref{sec:nnps}. These may be either set
    manually or be changed from the command-line.

  \item The \code{Solver.setup} method is then called which passes the created
    particles, the equations, NNPS, and kernel to the solver. The solver is
    constructed with the integrator information as shown in
    Listing~\ref{lst:ex:create_solver}. This method now has all the necessary
    information to generate the required code for the equations and the
    integrator. An \code{SPHCompiler} object is delegated the task of
    generating the code, compiling it, and being able to call it from Python.

  \item The application also sets up any user callback functions that need to
    be called after each step of iteration or after each stage of integration.

  \item Finally, the \code{Solver.solve} method is called which performs the
    time integration. The integration process periodically dumps output files
    as configured by the user.
  \end{enumerate}
\end{sloppypar}




The generation of the high-performance code from the Python equations and
integrators is an involved process. PySPH has two important classes to manage
the integration of the particle properties. The \code{Integrator} steps the
particle properties in time and computes the accelerations using another class
called the \code{AccelerationEval}. These Python classes internally use a
compiled module that implements all the high performance aspects of the
computation. On the CPU, the compiled code is in the form of a
Cython~\cite{Behnel:2011:CBB:1957373.1957467} extension module. Whereas, on the
GPU devices, these are implemented as functions provided by the underlying GPU
wrapper (PyOpenCL and PyCUDA~\cite{Klockner:2012:PPS:2109228.2109321}). We
discuss these in more detail in the following.

The \code{AccelerationEval} itself merely exposes a Python interface which can
be called. The \code{SPHCompiler} uses an \code{AccelerationEvalHelper} class
which generates the appropriate code and sets up the underlying
high-performance module. A similar strategy is employed for the
\code{Integrator}.

All the acceleration equations follow a very similar structure where the methods
of a group of equations are invoked for each particle. Taking advantage of the
similarity, we employ a technique commonly used in web development called
\emph{templating} to write the common code once and dynamically substitute the
variable parts. PySPH uses the mako~\cite{soft:mako} template library for this
and auto-generates the Cython code using a mako template.

To understand this better we consider the case of dam-break
problem. We have two groups involving two different particle arrays and five
different equations. We can break this up into pair-wise interactions between
the respective particle arrays as below:
\begin{sloppypar}
  \begin{enumerate}
  \item For all the fluid particles, evaluate the \code{EOS}.
  \item For all the solid particles, evaluate the \code{EOS}.
  \item For all the fluid particles (as destination), evaluate the
    \code{ContinuityEquation} and \code{MomentumEquation} due to the fluid
    particles.
  \item For all the fluid particles (as destination), evaluate the
    \code{ContinuityEquation} and \code{MomentumEquation} due to the solid
    particles.
  \item For all the solid particles (as destination), evaluate the
    \code{ContinuityEquation} due to the fluid particles.
  \end{enumerate}
\end{sloppypar}

Consider the continuity equation evaluated on the fluid by the fluid itself. The
pseudo-code (written with a Python syntax) to evaluate this would be as shown in
Listing~\ref{lst:eval-equation}. As can be seen in the listing, we first set the
initial value of the property by calling the method implemented by the user and
then call the loop with the appropriate neighbors. Clearly, executing this in
parallel would require a lot more work and computation of the neighbors is
itself fairly involved. However, the basic steps of the SPH computation are
essentially what is shown in the algorithm. We write a mako function that
generates the code necessary for each pair of particle arrays given a group of
equations. Depending on the combination of particle arrays and equations, this
is automatically expanded out into many lines of code. This is a very powerful
approach because the template is written only once.
\begin{lstlisting}[label={lst:eval-equation},caption={Pseudo-code for the
evaluation of a single equation.}]
for d_idx in range(n_dest_particles):
    continuity_fluid.initialize(d_idx, fluid.arho)

for d_idx in range(n_dest_particles):
    for s_idx in fluid.neighbors(d_idx):
        DWIJ = kernel.gradient(...)
        VIJ = [fluid.u[d_idx] - fluid.u[s_idx], ...]
        continuity_fluid.loop(
            d_idx, s_idx, fluid.arho, fluid.m, DWIJ,
            VIJ
        )
\end{lstlisting}
The key part of generating the high performance implementation of the equations
is delegated to the \code{compyle}
package~\cite{compyle-scipy-2020,soft:compyle}. Since Cython is very similar to
Python, we create equivalent Cython classes and determine the types of the
variables passed to the equations using the type information from the particle
arrays. The integrator is also similar and another mako template is used for
this. All of the generated code (which includes the high-performance classes for
equations and integrator steppers) is injected into a Cython file, automatically
compiled and linked, and called transparently by PySPH.

Adding support for execution via OpenMP requires rewriting the generated
acceleration evaluation to include OpenMP loops, managing issues like false
sharing carefully, and dealing with the special variables like \code{VIJ, DWIJ}
etc. However, all of this needs to be done just once and when implemented, the
code can be executed on multiple cores. It is important to note that the
user-facing code does not change.

On a GPU, this process is similar but requires a lot more effort. The first
issue is that on OpenCL or CUDA we are not able to create classes that are
analogous to the Cython ones. Instead we create GPU kernels for each function
and invoke them. The user code in the equations and integrator steps also must
be transpiled into OpenCL or CUDA. This is done by a special transpiler which
takes low-level Python code and converts it to efficient C code. The transpiler
is part of the \code{compyle} package. This allows PySPH to generate
high-performance GPU code on the fly. The rest of the process is similar
and uses mako templates to generate GPU code.

It is important to note that there are only 4 mako templates used in
PySPH. There are two for the CPU code generation (one for the integrator and one
for the acceleration evaluation) and two more for the GPU. Thus, by employing
mako templates in this fashion we are able to significantly reduce the
complexity of our implementation without losing a significant amount of
readability.

As discussed earlier, there are some severe restrictions in what code can be
written in the various equation methods. The Python programming language uses a
global interpreter lock (GIL) in order to ensure thread safety. Hence, any
computations involving multiple threads requires that no Python objects be
manipulated. The PySPH particle arrays expose low-level memory views of the
underlying properties and constants. The equations are therefore written to only
manipulate these properties and constants.  This can be done entirely in
parallel and is one of the most time-consuming parts of an SPH simulation. Thus,
we do not encourage users to write any arbitrary Python code in the equation
methods. The \code{compyle} package also follows the same philosophy and users
write a restricted subset of Python where only manipulation of one dimensional,
contiguous numpy arrays are allowed. Memory allocation is not allowed and Python
objects may not be created. More details on the restrictions are provided in
\cite{compyle-scipy-2020}. Compyle provides important parallel algorithms like
the elementwise operations, reductions, and scans. PySPH uses \code{compyle} for
code generation, transpilation, and some of the parallel algorithms.

Thus far the general features and functioning of the high-level framework have
been discussed. It is important to note that it is possible to use PySPH as a
library too. Much of the functionality is usable outside of the context of the
framework. For illustrative purposes, let us consider evaluating the continuity
and momentum equations mentioned earlier on a particle array this can be done
using the code shown in listing~\ref{lst:sph-eval}. Here, the
\code{SPHEvaluator} can be used to evaluate the effect of the equations on the
particles. This allow a user to reuse the infrastructure to perform a variety
of computations with particles and their nearest neighbors.
\begin{lstlisting}[label={lst:sph-eval},caption={Code to evaluate a collection
of equations.}]
from pysph.base.utils import get_particle_array
from pysph.base.kernels import QuinticSpline
from pysph.tools.sph_evaluator import SPHEvaluator

pa = get_particle_array(name='fluid', x=x_data, y=y_data, ...)
eqs = [Group(equations=[
    ContinuityEquation(dest='fluid', sources=['fluid']),
    MomentumEquation(dest='fluid', sources=['fluid'])]
)]
evaluator = SPHEvaluator(
    arrays=[pa], equations=eqs, dim=2, kernel=QuinticSpline(dim=2)
)
evaluator.evaluate()
\end{lstlisting}

To summarize the overall design, PySPH uses a high-level \code{Application}
class to abstract out the basic pieces involved in an SPH computation. Python
is used to create the particles using numpy arrays, specify the particle
interactions, and integration process. This information is then extracted in a
systematic way and high-performance code is generated as necessary to compute
the particle interactions on different HPC hardware.

\subsection{Advanced features}\label{sec:advanced}

PySPH also supports some advanced features that are useful in certain
contexts. In this section we discuss two such features, iteration of a group
of equations while evaluating the RHS and inlet and outlet boundary
conditions. These features are used in some of the results presented.

\subsubsection{Iterated groups}\label{sec:iterated_group}

In SPH, some schemes like ISPH, pressure is determined using pressure Poisson
equation (PPE), which is solved iteratively. In order to evaluate a group of
equations `iteratively` in between other groups which are being solved once per
time step, we have implemented iterated~\code{Group}s in PySPH.

Iterated~\code{Group}s are~\code{Group}s as explained in
Section~\ref{sec:design-example} with an additional feature to iterate within a
time step. In Listing~\ref{lst:iterated-groups}, we show the set of equations
which are to be iterated in order to get a converged solution. The
equations~\code{PressureCoeffmatrixiterative} and~\code{PPESolve} (\code{dest}
and~\code{sources} are not shown) are grouped together and put in an iteration
group as indicated by the~\code{iterate} keyword. One may also specify the
minimum and maximum number of iterations required. The equation~\code{PPESolve}
overloads a method called~\code{converged}.  This \code{converged} function
returns $-1$ when the iterations have not converged and $+1$ otherwise. The
entire group of equations is re-executed until all equations have converged or
if the maximum number of iterations is reached. The convergence criterion is
dependent on the algorithm chosen. For example, if one wishes to ensure that the
average change to the pressure in the entire simulation is less than a
prescribed tolerance, then one can continue iteration until this criterion is
attained by returning a suitable value.
\begin{lstlisting}[label={lst:iterated-groups},caption={Code showing the usage
and implementation of iterated~\code{Group} in PySPH.}]
 Group(equations=[
        Group(equations=[
            PressureCoeffMatrixIterative(...),
            PPESolve(...)
        ])
    ], iterate=True, min_iterations=2,
    max_iterations=1000)
\end{lstlisting}
In~\cite{muta2019simple} we implement a simple iterated scheme for
incompressible SPH where a system of equations is solved using Jacobi iterations.
This is done using an iterated group.

\subsubsection{Inlet/outlet boundary conditions}\label{sec:io}

The imposition of inlet/outlet boundary condition is non-trivial in SPH. The
method to implement differs from scheme to scheme. In~\cite{negi2019improved},
we explored various type of inlet/outlet boundaries. An inlet/outlet boundary
generally requires ghost particles to obtain the properties from the
fluid. PySPH provides the equations which are to be used for interpolation, the
stepper to move the particles as explained in Section.~\ref{sec:design-example},
and a callback function to convert the inlet/outlet particles to fluids or
vice-versa.

In PySPH, the class \code{InletOutletManager} (IOM) manages all the work
needed for a inlet/outlet boundary. In order to create an
\code{InletOutletManager} object, we need information of all inlets and
outlets in the form of start/end coordinates, outward normal, additional
equations etc. The IOM performs the following tasks:
\begin{enumerate}
\item Create a callback which adds or deletes the particles for fluid to
  inlet/outlet.
\item Create ghost particles.
\item Pass stepper functions to the solver.
\item Provide interpolation equations to the scheme.
\end{enumerate}
Currently, PySPH implements all the inlets/outlets described in
\citet{negi2019improved}.

\subsection{Distributed execution}

As discussed earlier, one may use MPI to run PySPH applications in a distributed
mode. This works by making use of another sister package called
\code{PyZoltan}~\cite{soft:pyzoltan}, which is a wrapper to the ZOLTAN
package~\cite{zoltan-web,zoltan:2012}. PySPH uses PyZoltan to partition the
particles across multiple computers. A ghost region is defined over each
processor and any particle in these regions are shared between appropriate
processors. PySPH implements the necessary parts that allow ZOLTAN to distribute
and gather the particles. This is all managed by a \code{ParallelManager}
available in the \code{pysph.parallel} package. The parallel manager is invoked
each time the particles have moved and the properties are changed.  Updated
properties of particles in the ghost region are exchanged with other processors
as required. Any particles that have moved outside the region of a given process
are shifted to the appropriate processor. PyZoltan also manages the load
balancing so the particles are evenly distributed. More details are available in
\citet{kunal-parcomptech-2013}.

\subsection{Reproducibility}

PySPH aims to be a research tool that facilitates reproducible computational
science. The first important feature in this context is that PySPH examples
ship with the software and are designed to be reusable. When PySPH is
installed, an executable \code{pysph} is made available and this can be used
to execute any of the shipped examples. There are around 90 examples at
various levels of maturity. Many of the standard benchmark problems seen in
the literature are available and these also include post-processing of the
results along with comparisons with exact or experimental results in some
cases. For example, if we execute
\begin{verbatim}
$ pysph run taylor_green --help
\end{verbatim}
We will see that we can execute this standard problem with 7 different SPH
schemes that are already implemented in PySPH. The scheme can be chosen by using
the \code{--scheme scheme-name} command line argument. When the problem finishes
executing, the results are also compared with the exact solution. Many of these
examples support command line arguments that make it possible to perform
parametric studies.

\begin{sloppypar}
  Furthermore, as the examples are part of PySPH, a user may easily subclass
  the example to modify it with a minimum of code replication by importing the
  example class as \code{from pysph.examples.taylor_green import TaylorGreen}.
\end{sloppypar}

In addition to this, PySPH integrates with a sister package called
automan~\cite{pr:automan:2018}. The present work and our recent papers in the
area of SPH are
reproducible~\cite{edac-sph:cf:2019,muta2019simple,negi2019improved,pr:dtsph:2019}.
What we mean by this is that every figure produced in these papers is fully
automated. We provide a link to the repository so anyone may reproduce all the
computations (or a part of them) by running a single command. The advantages of
this to the readers and researchers are clear. We have also found that investing
time in automation is also very valuable to the authors. This is discussed in
detail in \cite{pr:automan:2018}.

Thus, PySPH is a powerful framework for reproducible research in particle
methods in general and SPH in particular. Many of the tools created for PySPH
may be used in other contexts.

\subsection{Software engineering}

PySPH follows some important and useful software engineering practices to
improve the quality of the code.
\begin{itemize}
\item \textbf{Open code and repository}: Our code is open source and ships
  with a liberal BSD 3-clause license. While the code is copyrighted, it may
  be reused by anyone for any purpose and may also be redistributed. The code
  is openly available on \url{https://github.com/pypr/pysph} and does not
  require any formal process to access or download. \code{git} is used for
  version control.
\item \textbf{Testing}: We have implemented a large number of unit tests and
  functional tests that ensure correctness of our code to the extent
  possible. The test coverage can be further improved, however, we already have
  close to 900 tests. These tests are executed on Linux and Windows
  computers. Links to the test results are available on the project page. The
  tests include checking that the kernels are accurate and satisfy the various
  integral properties, correctness of the nearest neighbor computations for the
  different methods and for the CPU and GPU backends, correctness of the various
  methods of the particle array. The parallel execution with OpenMP and MPI is
  tested as compared to a serial execution for some time steps for a few
  problems in two and three dimensions. The acceleration evaluation is tested
  for simple equations that exercise all of the various features on all the
  supported backends, the integrators are also tested on simple problems. All of
  the examples that are included are compiled and executed for a single time
  step.  We do not test the specific schemes as that would be very time
  consuming.  The current set of tests take about half an hour to execute on
  Travis-CI and about 15 to 20 minutes on appveyor. When a bug is found a
  suitable test case is introduced.
\item \textbf{Continuous Integration}(CI): PySPH employs continuous integration
  testing. New features and bug fixes are added in the form of pull-requests.
  When these pull-requests are submitted, all the tests are executed on Linux
  and Windows machines automatically. This immediately ensures that any
  existing functionality is not broken by the new feature. This is standard
  practice for most popular open source software.
\item \textbf{Online documentation}: The PySPH documentation, which is
  extensive, is automatically built every time the master branch of the code
  repository is changed. The documentation is hosted at
  \url{https://pysph.readthedocs.io} and is available for anyone to read.

\item \textbf{Support}: Bug reports and issues may be filed on the github web
  interface by any user. These are attended to by the PySPH developers as time
  permits. A mailing list is also available at
  \url{https://groups.google.com/d/forum/pysph-users} where users may ask
  questions and also see any PySPH related announcements.
\end{itemize}
We believe that by using well established software engineering practices,
PySPH will be easy to use, well supported, and retain good quality over
time.


\section{Sample results}
\label{sec:results}

In this section, we demonstrate the capabilities of PySPH by solving a wide
variety of problems in different fields, including incompressible and
compressible fluid dynamics, and elastic dynamics. At the end of this section
we provide qualitative indicators analysing the performance of PySPH on CPU
(both single and multi-cores) and GPU architectures. The results presented
here are only indicative of the breadth of problems that can be handled. We do
not provide all the details of the parameters used in the simulations.
However, the source code for the simulations presented are available at the
repository \url{https://gitlab.com/pypr/pysph_paper}.

\subsection{Taylor-Green Vortex}

\begin{figure}[!h] \centering
  \begin{subfigure}{0.45\linewidth}
    \includegraphics[width=1.0\linewidth]{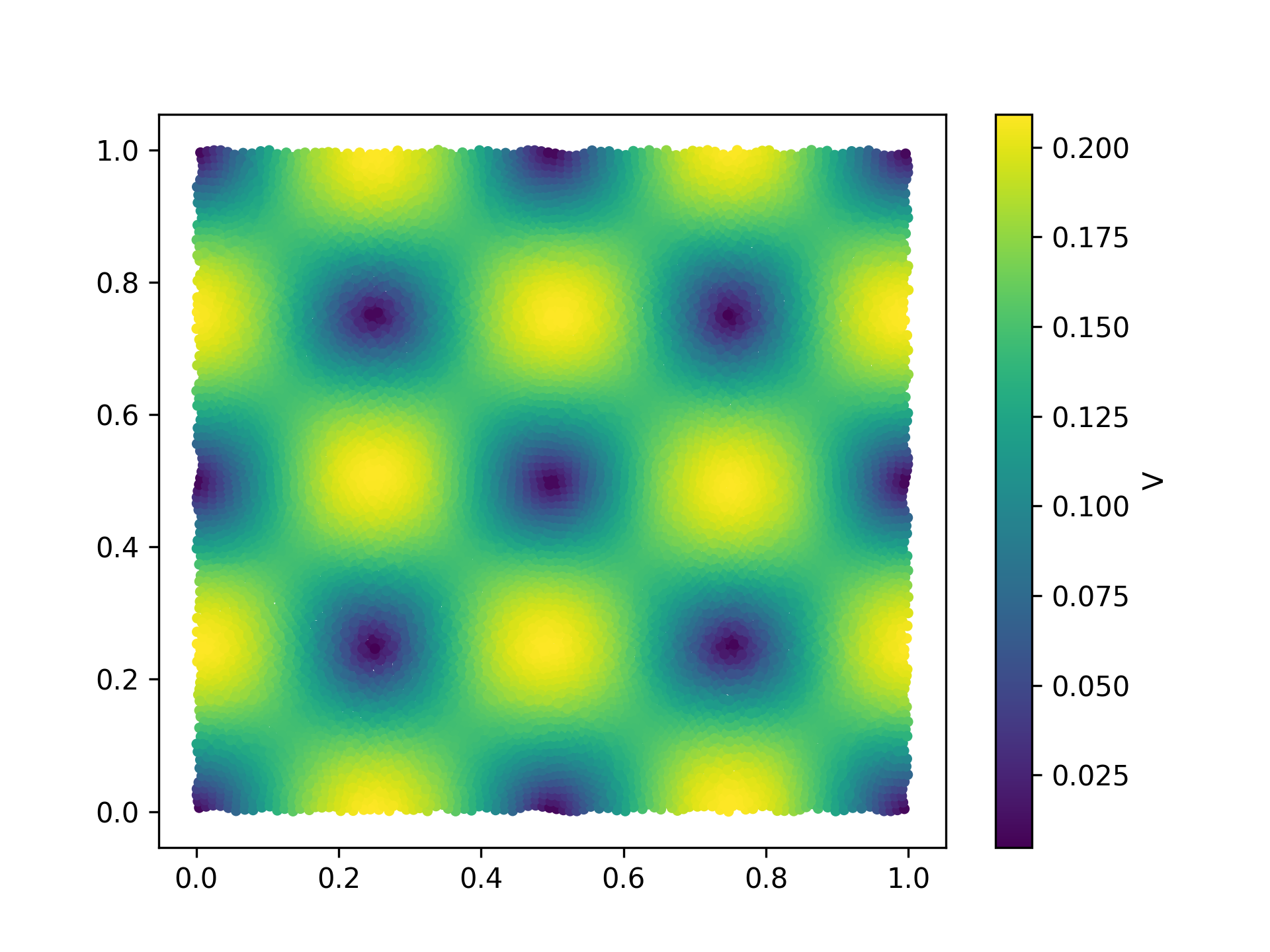}
  \end{subfigure}
  \begin{subfigure}{0.45\linewidth}
    \includegraphics[width=1.0\linewidth]{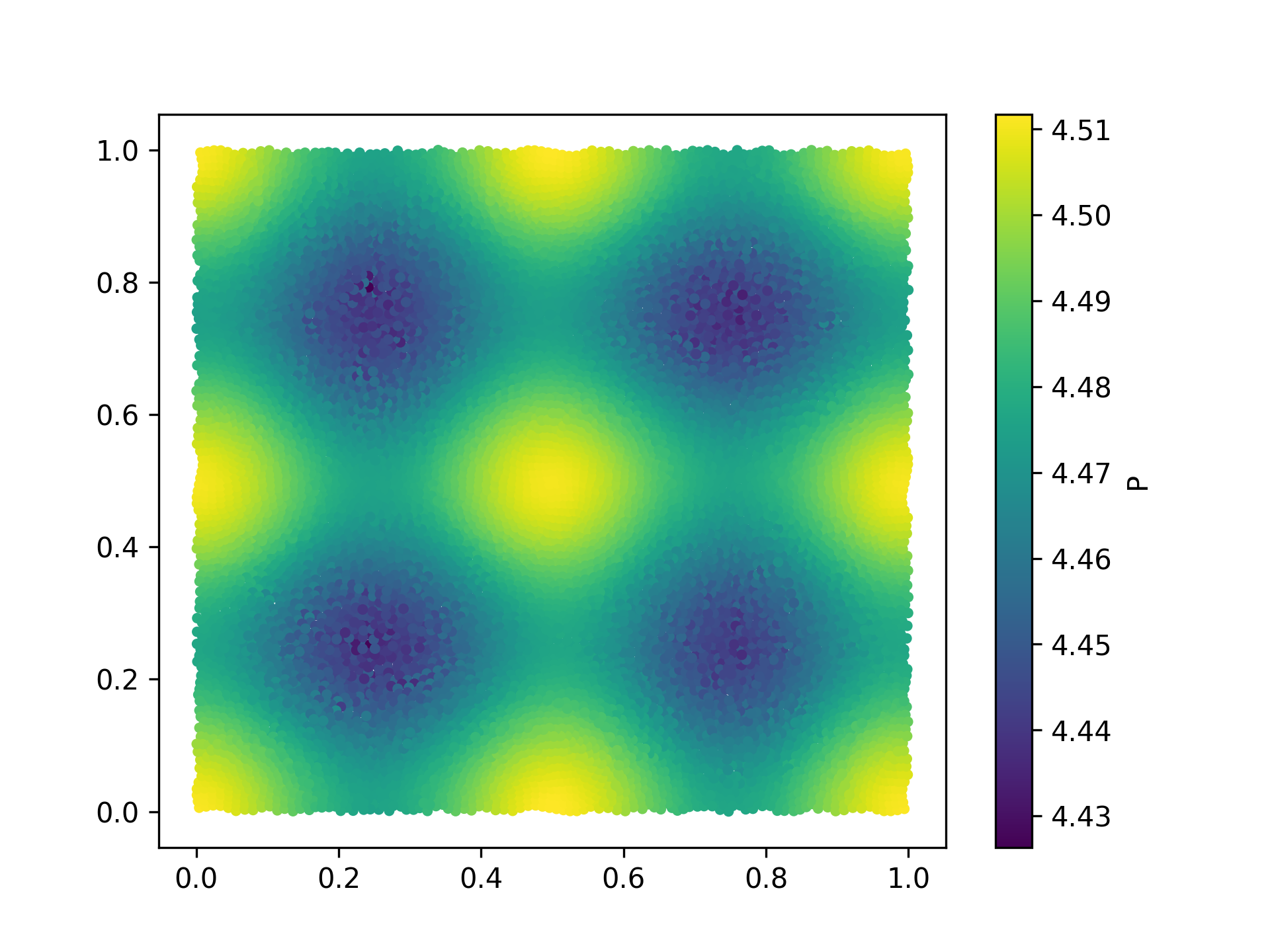}
  \end{subfigure}
\\
\caption{Velocity magnitude, and pressure distribution for Taylor-Green
  vortex at $t=2$s using EDAC~\cite{edac-sph:cf:2019}.}\label{fig:tg}
\end{figure}

\begin{figure}[h!]
  \centering
  \includegraphics[width=0.5\textwidth]{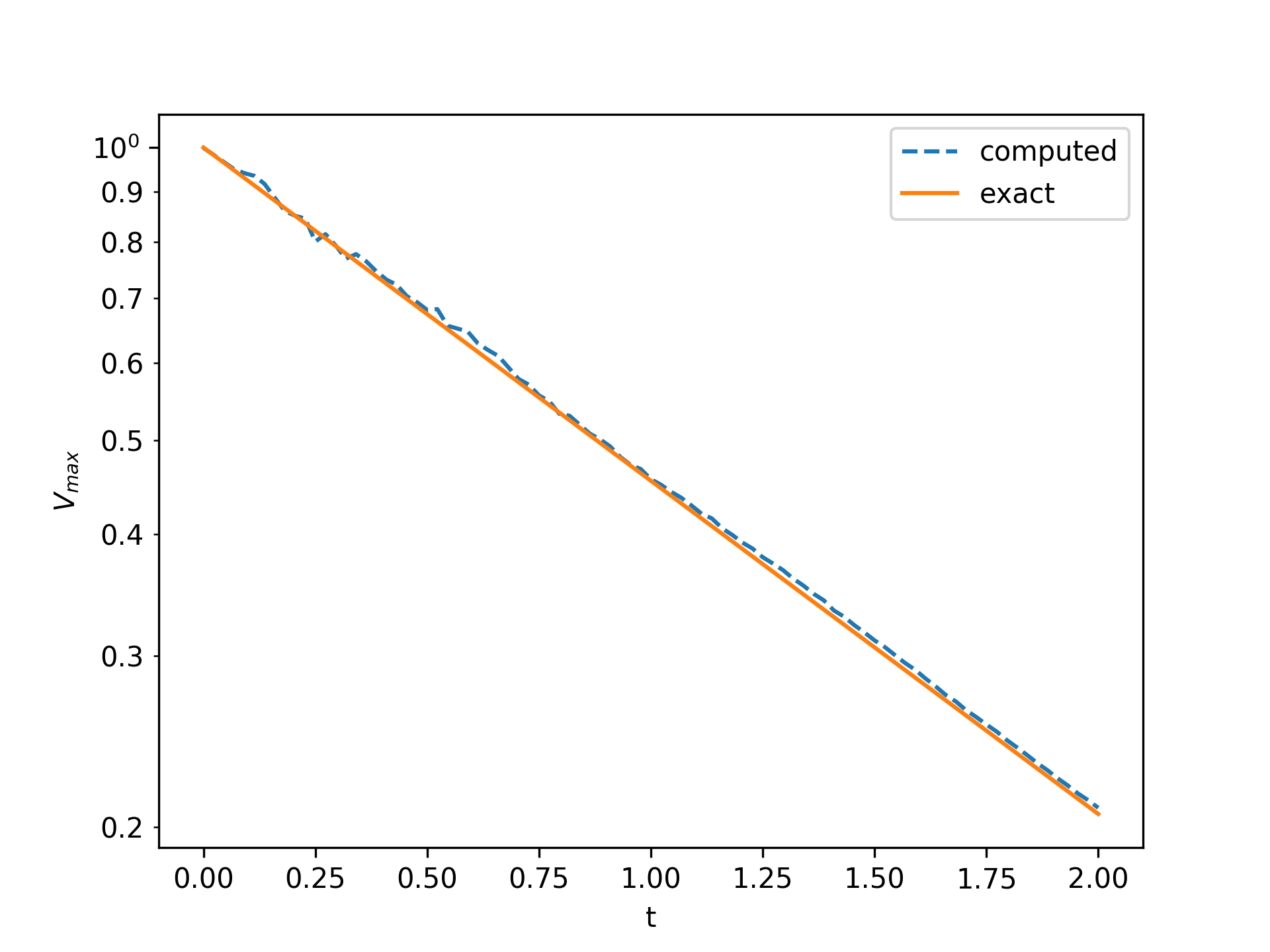}
  \caption{Maximum velocity decay in Taylor-Green vortex compared with the exact
    solution.}\label{fig:decay}
\end{figure}

Taylor-Green vortex is a standard test case for incompressible flows. This
problem is periodic in both $x$ and $y$ directions and has no boundaries. This
problem admits an exact solution given by,
\begin{equation*}
\begin{split}
  u & = -U e^{bt} \cos(2 \pi x) \sin(2 \pi y), \\
  v & = U e^{bt} \sin(2\pi x) \cos(2\pi y), \\
  p & = -U^2 e^{2bt} (\cos(4\pi x) + \cos(4\pi y))/4,  \\
\end{split}
\end{equation*}
where $U$ is the maximum velocity, $b=-8 \pi^2 / Re$, and $Re$ is the Reynolds
number. In order to simulate this problem, we consider a periodic domain of
size $1m \times 1m$ and use the Entropically Damped Artificial Compressible
SPH (EDAC-SPH) scheme~\cite{edac-sph:cf:2019}. The maximum velocity, $U$ and
Reynolds number, $Re$ are set to $1$m/s and $100$ respectively. The domain is
discretized by $100\times 100$ particles, and $h/\Delta x = 1.2$, where
$\Delta x$ is the particle spacing. In Fig.~\ref{fig:tg}, we plot the velocity
and pressure at $t=2$ seconds. The initial pattern is maintained, with the
decay in velocities as expected. In Fig.~\ref{fig:decay}, we plot the maximum
velocity compared with the expected theoretical result and the results are in
agreement.

\subsection{Lid-driven cavity}
\begin{figure}[h!]
  \centering
  \includegraphics[width=0.5\textwidth]{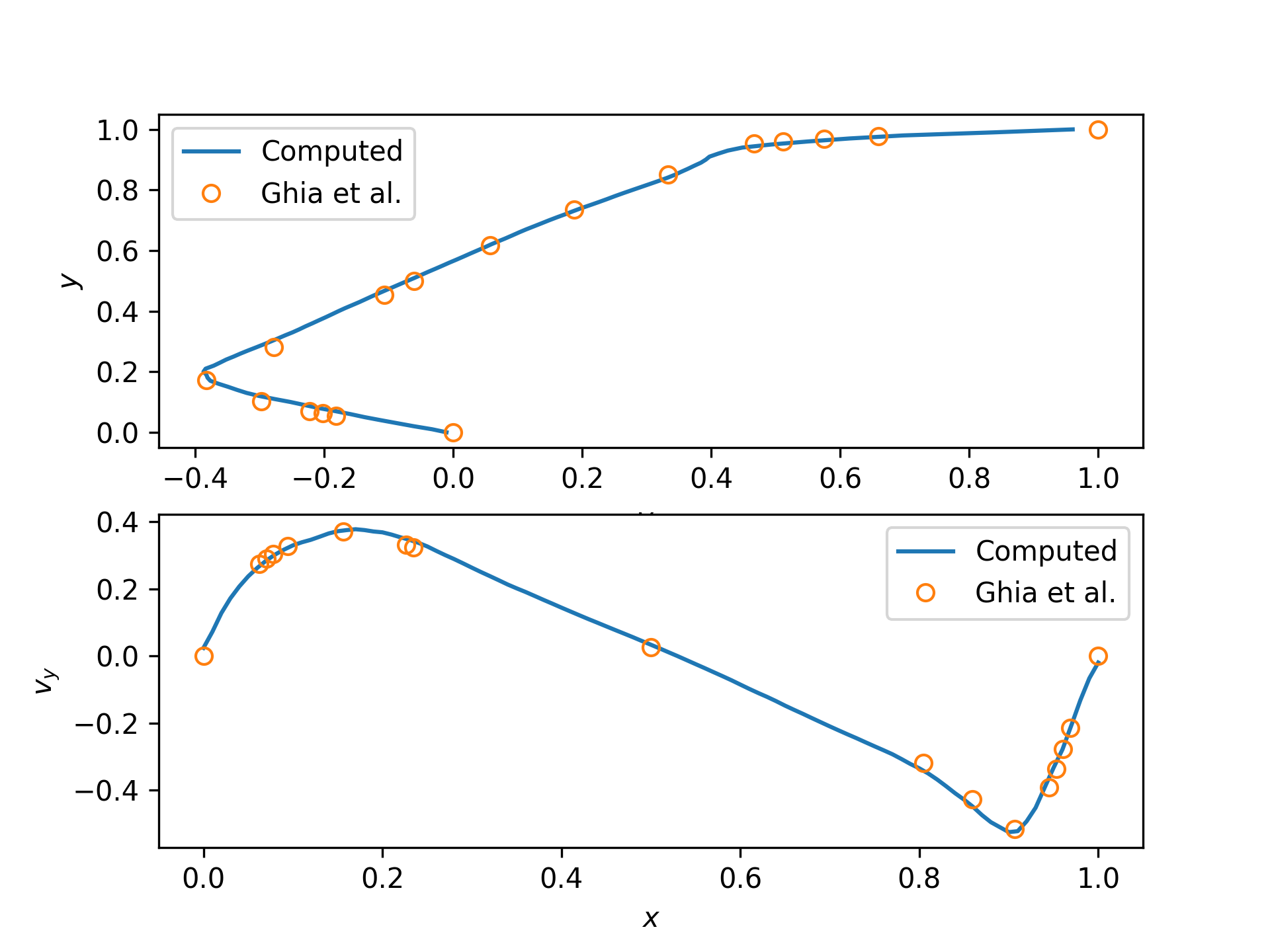}
  \caption{Velocity profiles $u$ vs. $y$ and $v$ vs. $x$ for the lid-driven-cavity
    problem at Re = 1000. Particle discretization of $100 \times 100$. We
    compare the results with those of~\citet{ldc:ghia-1982}.}\label{fig:ldc}
\end{figure}

The lid-driven cavity is a classic benchmark problem. We consider a 2D cavity
of size $1m\times 1m$ with the top boundary moving with a velocity $U=1$m/s
and use the Transport Velocity Formulation (TVF)~\cite{Adami2013} scheme. The
Reynolds number, $Re=1000$ is considered. $50 \times 50$ particles are used to
discretize the problem and $h/\Delta x = 1.0$. The quintic spline kernel is
used. In Fig.~\ref{fig:ldc}, we compare the velocity profile at the centerline
with \citet{ldc:ghia-1982}. The results are in agreement showing the validity
of the TVF scheme for this problem.

\subsection{Flow past a bluff body}
Here we simulate the flow past a circular cylinder in an internal flow. This
test case is used to demonstrate the implementation of inlet outlet boundary
conditions in SPH. The inlet velocity is set to $1$m/s, the diameter of the
cylinder is $2$m, and the Reynolds number is 200. The domain is discretized
such that there are $20$ particles along the diameter of the cylinder. The
inlet continuously feeds the particles into the fluid domain with desired
initial properties, while the outlet removes the fluid particles smoothly from
the flow. Further details of the theory and implementation can be found
at~\cite{negi2019improved}.

\begin{figure}[h!]
  \centering
  \includegraphics[width=\textwidth]{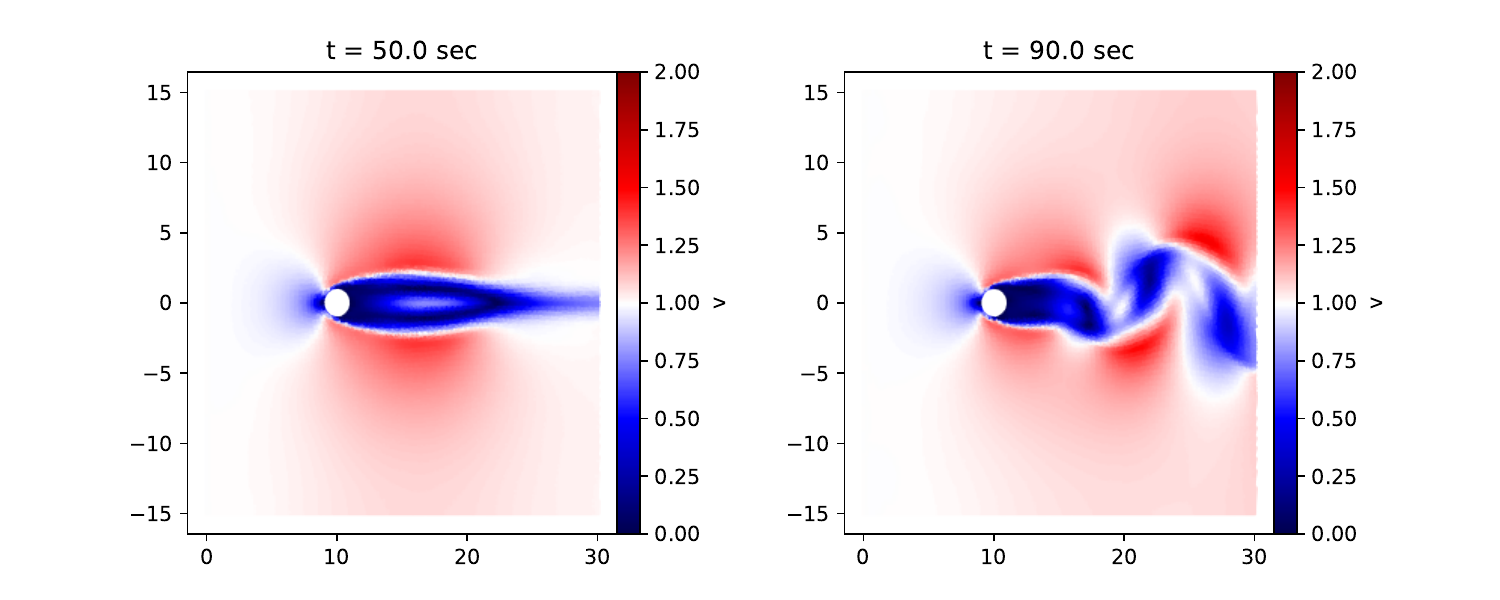}
  \caption{The plots showing the velocity magnitude of the flow past a
  circular cylinder at $t = 50.0$s and at $t = 90.0$s.}\label{fig:fps}
\end{figure}

Fig.~\ref{fig:fps} shows the velocity magnitude of the flow past the cylinder.
The vortex shedding at time $t = 90$s is seen clearly. As a quantitative
validation we determine the lift and drag coefficients for the cylinder, which
have been found to be $1.524$ and $0.722$. There is a 5\% difference from
those reported by~\cite{tafuni2018}. The Strouhal frequency is found as 0.2
and this differs from other reported results by 2\%.

\subsection{Dam break in 3D}

\begin{figure}[!h] \centering
  \begin{subfigure}{0.32\linewidth}
    \includegraphics[width=1.0\linewidth]{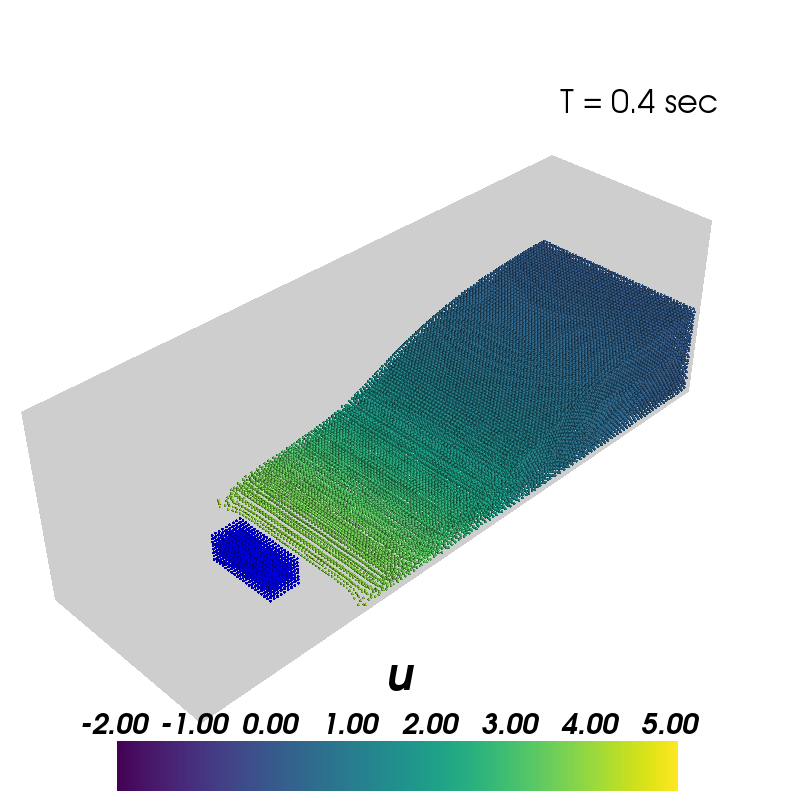}
  \end{subfigure}
  \begin{subfigure}{0.32\linewidth}
    \includegraphics[width=1.0\linewidth]{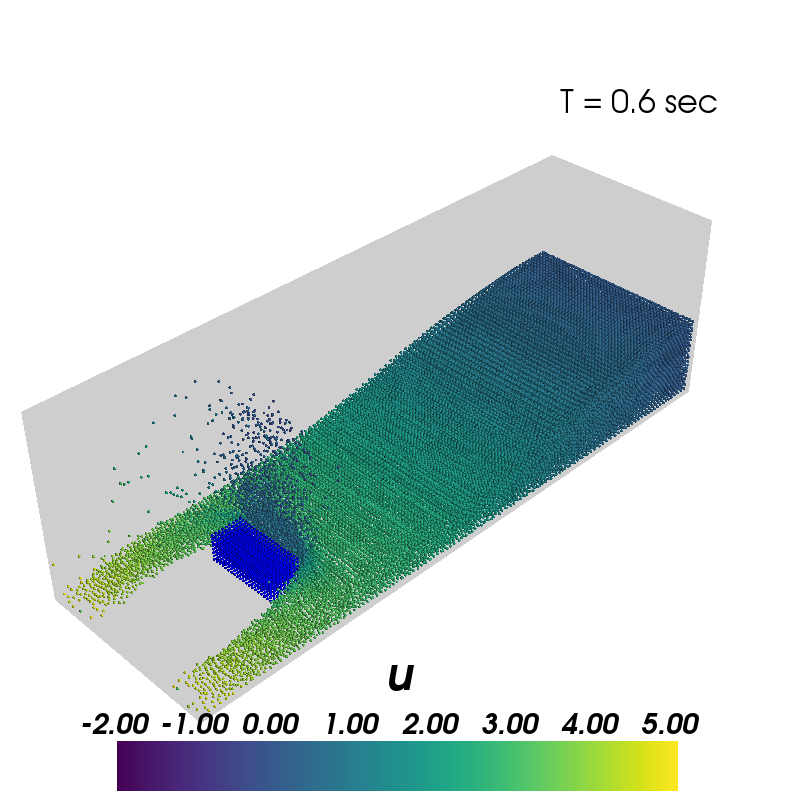}
  \end{subfigure}
\\
  \begin{subfigure}{0.32\linewidth}
    \includegraphics[width=1.0\linewidth]{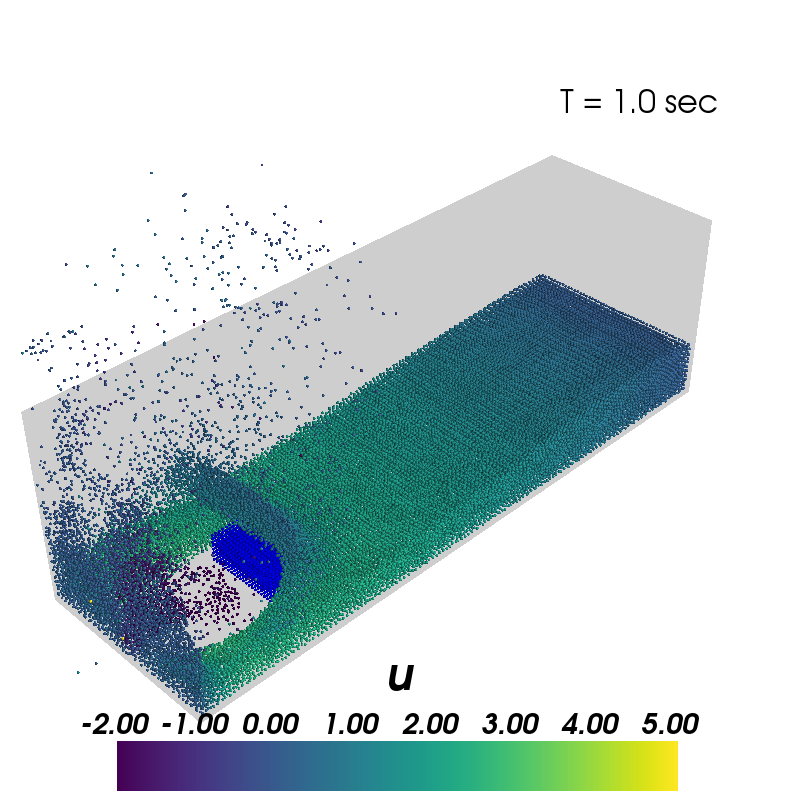}
  \end{subfigure}
  \begin{subfigure}{0.32\linewidth}
    \includegraphics[width=1.0\linewidth]{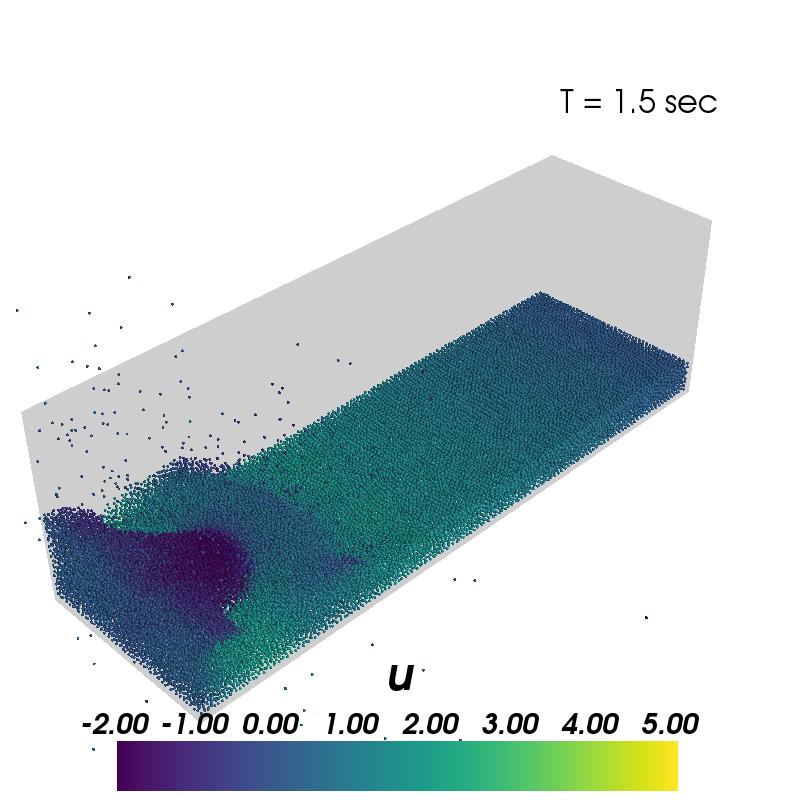}
  \end{subfigure}
  \caption{Figure showing the velocity magnitude of the particles in the three
    dimensional dam break simulation at various time instances.}\label{fig:db3d}
\end{figure}

A three-dimensional dam break is simulated using a simple iterative ISPH
(SISPH) scheme which is discussed in greater detail in~\citet{muta2019simple}.
This is an iterative incompressible scheme which uses the iterated \code{Group}
feature as discussed in Section~\ref{sec:iterated_group}. A fluid column is
released from one side of a tank under the influence of gravity, an obstacle is
placed in the path of the fluid flow. In Fig.~\ref{fig:db3d}, we show the
$x$-component of the velocity at different time steps. As can be seen, the flow
splashes at the obstacle.

\subsection{Kelvin-Helmholtz instability}

\begin{figure}[!h]
  \centering
  \includegraphics[width=0.6\textwidth]{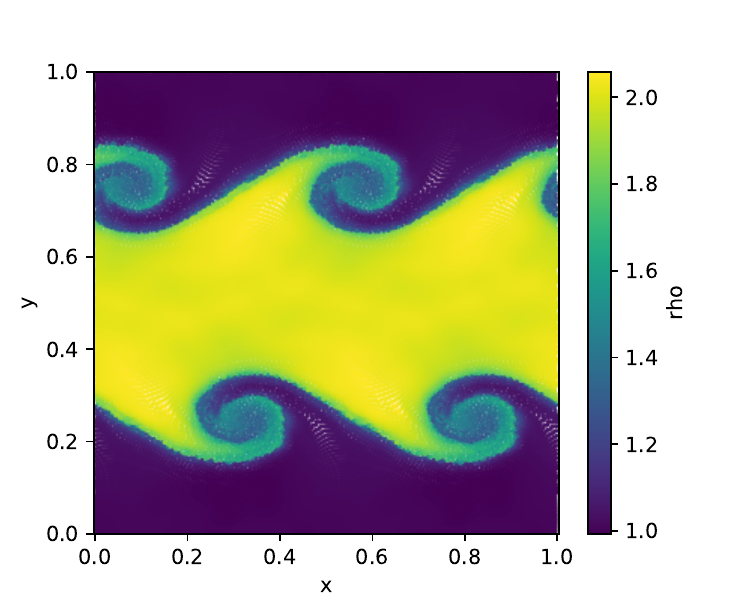}
  \caption{Kelvin-Helmholtz instability simulation using
    CRKSPH~\cite{crksph:jcp:2017} with $200 \times 200$ particles. Shown is the
    density $\rho$ of the particles at $t = 2$s.}\label{fig:kh}
\end{figure}

This example demonstrates the modeling of fluid instabilities and compressible
flows in PySPH. We model Kelvin-Helmholtz instability, which exhibits the
subsonic instabilities of a compressible gas. Kelvin-Helmholtz instability
manifests at the interface of two fluids moving with two different velocity
profiles. The problem is periodic in both $x$ and $y$ directions. A sinusoidal
velocity perturbation is given in the $y$-direction as given
in~\citet{Robertson:2010}. This is simulated using the conservative
reproducing kernel SPH (CRKSPH)~\cite{crksph:jcp:2017} that is available in
PySPH. The density profile of the particles is shown in Fig.~\ref{fig:kh}. We
expect to see the roll-up of the shear layer, which is captured well by the
scheme.

\subsection{Sod's shock tube}

\begin{figure}[!h] \centering
  \begin{subfigure}{1.0\linewidth}
    \includegraphics[width=1.0\linewidth]{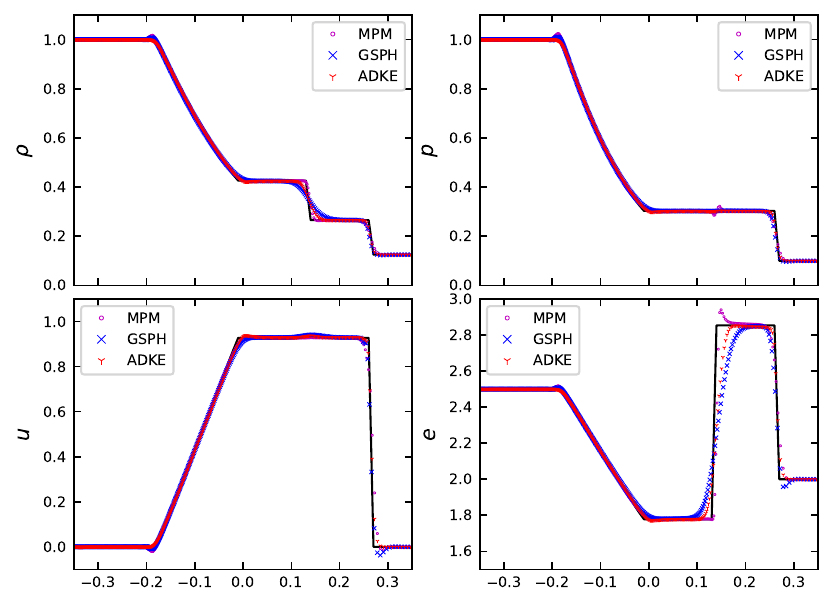}
  \end{subfigure}
  \caption{1D Sod's shock tube at $t=0.1$. Three schemes, ADKE, MPM and GSPH are
    used for the simulation. Comparison of the exact solution vs profiles of
    density, pressure, velocity, and energy of all the particles is shown.
  }\label{fig:shocktube}
\end{figure}

Sod's shock tube problem, considered a classic 1D problem in compressible fluid
dynamics. PySPH has several different compressible SPH schemes to solve such
problems. Here we simulate the problem using three schemes: Adaptive Density
Kernel Estimation technique (ADKE)~\cite{sigalotti2006}, Monaghan-Price-Morris
(MPM) scheme~\cite{price_2012}, Gudunov type Smoothed Particle Hydrodynamics
(GSPH)~\cite{puri2014}. For the simulation we use $720$ particles of which $640$
are placed with uniform spacing in the domain $x \in [-0.5, 0)$ and $80$ are
placed with uniform spacing in the domain $x \in [0, 0.5]$.
$(\rho_l,p_l,v_l)=(1.0,1.0,0.0)$ and $(\rho_r,p_r,u_r)=(0.125,0.1,0)$ on the
left and right sides of the diaphragm which is placed at $x = 0$. The results of
the simulation are compared with the exact solution in
Fig.~\ref{fig:shocktube}. For a detailed discussion of this problem, see
\cite{puri2014}.

\subsection{Elastic deformation of colliding rings}
\begin{figure}[h!]
  \centering
  \includegraphics[width=\textwidth]{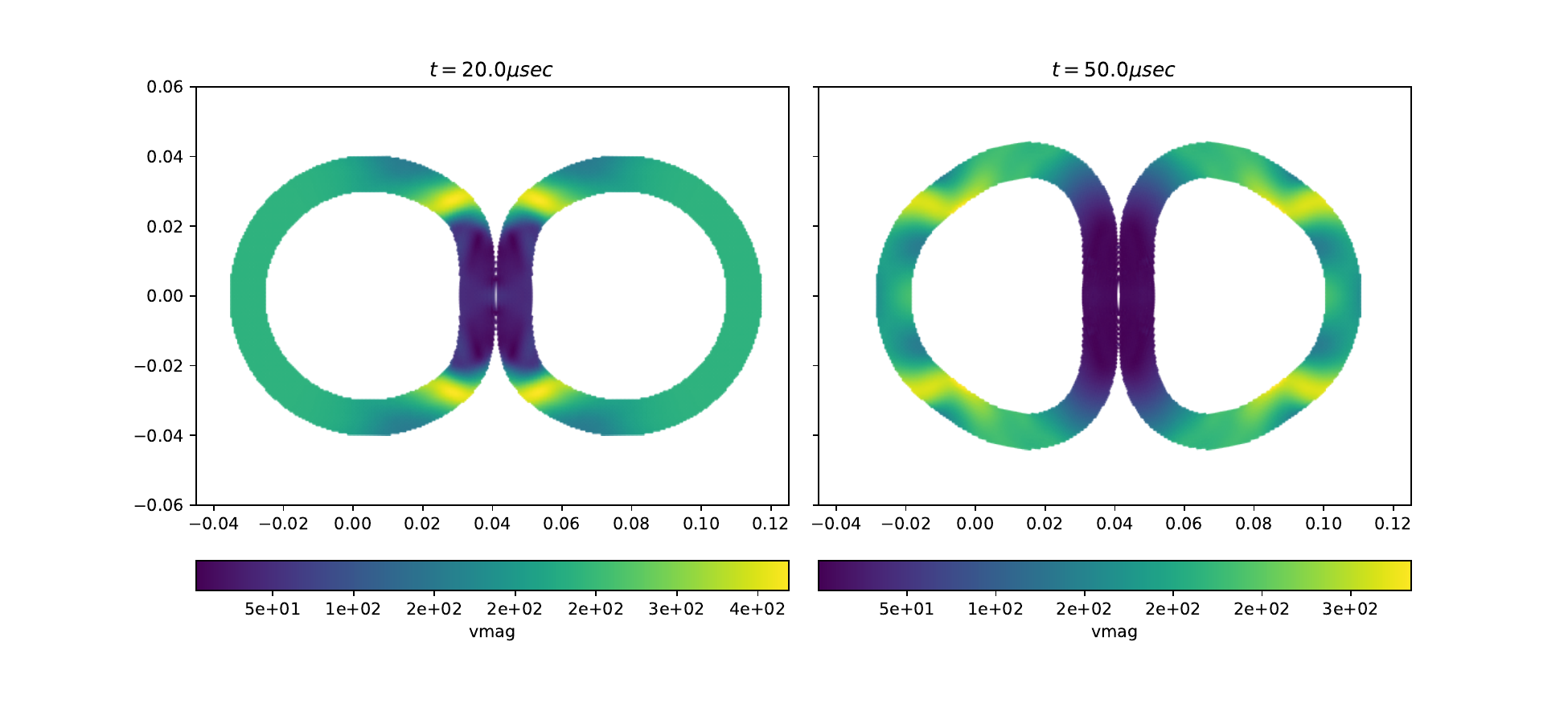}
  \caption{Velocity magnitude plots of the collision of elastic rings at times $t =
    20\mu$s and $t = 50\mu$s.}\label{fig:rings}
\end{figure}

In this example we simulate collision of two rubber rings. This example is used
to demonstrate the ability of PySPH in modelling elastic dynamics. We have
implemented the formulation proposed by~\citet{gray2001} to model the
elastic dynamics. Two rings are initialized with a velocity such that they are
moving towards each other. Figure~\ref{fig:rings} shows the velocity magnitude
at two different times. At $t = 20\mu$s the rings have collided with each other
and start to deform. At $50\mu$s, more significant deformation is seen,
eventually the two rings reverse their velocities although this is not shown in
the figure.

\subsection{Sand in a rotating drum}

\begin{figure}[h!]
  \centering
  \includegraphics[width=\textwidth]{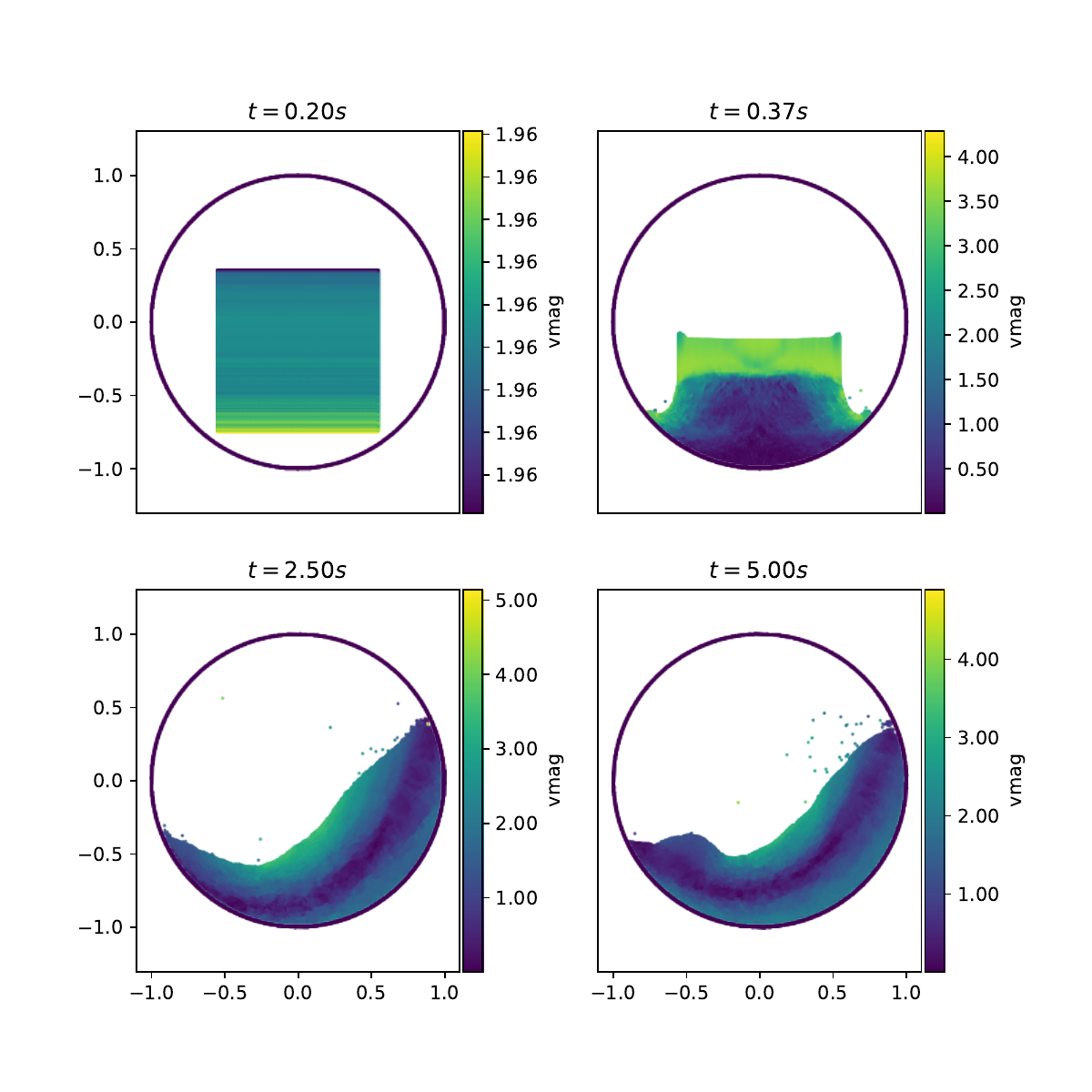}
  \caption{Positions of the sand in a drum, color indicates the velocity
    magnitude. The top row depicts the results when the drum is not rotating,
    and the bottom row is when the drum is rotating.}\label{fig:hopper}
\end{figure}

This example demonstrates an implementation of the Discrete Element Method (DEM)
implemented using PySPH. The motion of sand in a rotating drum is simulated.
The dynamics of the particles follow the formulation by Cundall and
Strack~\cite{cundall1979discrete} and Luding~\cite{luding2008introduction}. The
drum is initially static and starts rotating with an angular velocity of 5 rad/s
after the sand settles down. Fig.~\ref{fig:hopper} shows the arrangement of the
sand particles, the top row depicts the results when the drum is not rotating,
while the bottom row is when the drum is rotating. This shows that PySPH
provides the features required to implement a variety of different meshless
methods.

\subsection{Performance analysis}

\begin{figure}[!h]
  \centering
  \includegraphics[width=0.6\textwidth]{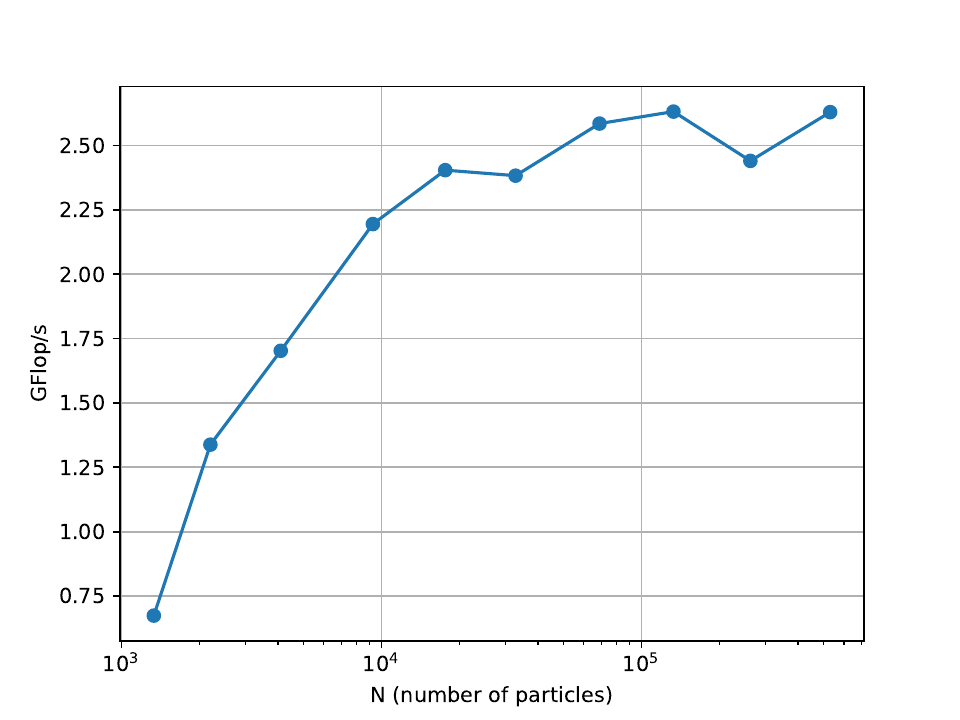}
  \caption{FLOPS vs.\ number of particles for a cube shaped block of water in
  free-fall on an i3-6100 CPU.}\label{fig:cpu_flops}
\end{figure}

In this section we discuss the performance of PySPH. Fig.~\ref{fig:cpu_flops}
shows the number of floating point operations per second (FLOPS) for a simple
simulation of a cube-shaped block of water falling in free-space under the
influence of gravity run on an i3-6100 CPU (single-core) while varying the
number of particles. We note that we compute the fluid dynamic equations on
the water particles during this computation. We use the Performance
Application Programming Interface (PAPI)~\cite{PAPI:2009} to count the FLOPS
using the CPU hardware counters.

\begin{figure}[!h]
  \centering
  \includegraphics[width=0.6\textwidth]{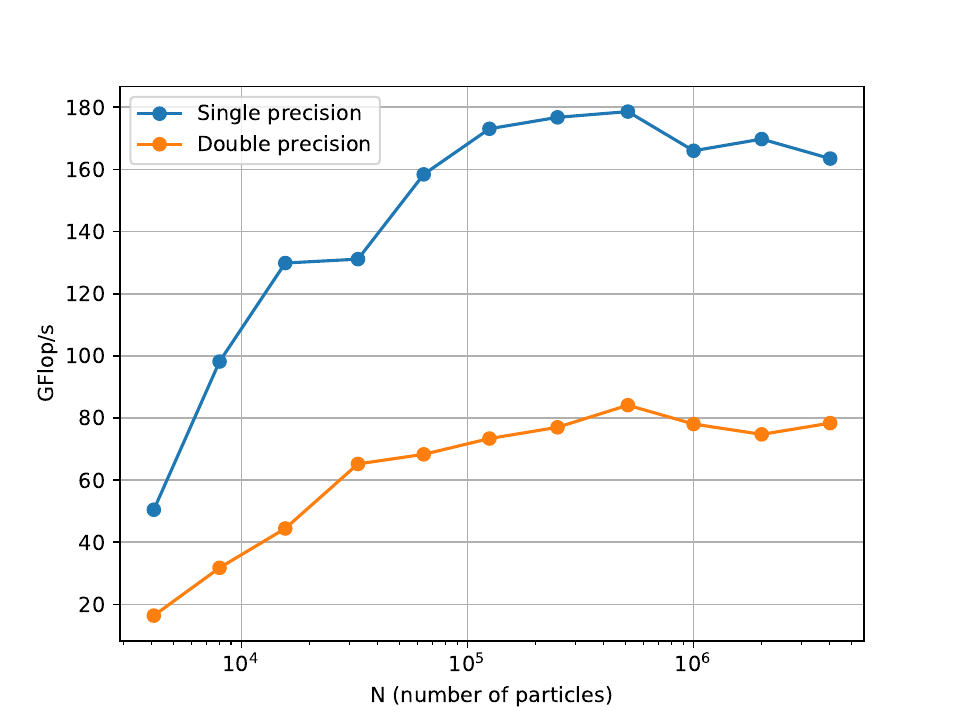}
  \caption{FLOPS vs.\ number of particles for the \code{loop} kernel of a cube
    shaped block of water in free-fall using single and double precision on a
    Tesla T4 GPU.}\label{fig:gpu_flops}
\end{figure}

\begin{figure}[!h]
  \centering
  \includegraphics[width=0.6\textwidth]{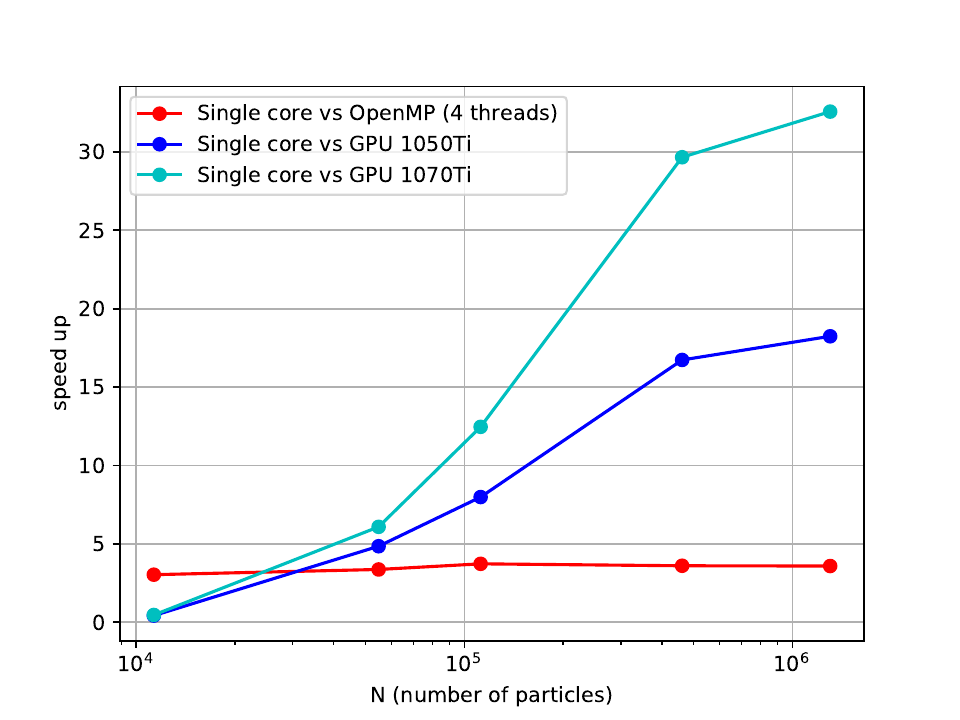}
  \caption{The comparison of number of particles vs.\ speed-up on different
    platforms with respect to the time taken by single core CPU for the three
    dimensional dam break.}\label{fig:speedup}
\end{figure}

We see that on the CPU, the performance peaks at about $2.6$ GFlop/s. Our
performance compares favorably with the singular value decomposition of an
$800 \times 800$ matrix using \code{numpy.linalg.svd} which uses the Intel
MKL~\cite{mkl_2014} implementation and is at $2.49$ GFlop/s that reduces as the
matrix size increases. It is also comparable to a 1D discrete Fourier transform
using \code{numpy.fft.fft} which also uses the MKL implementation and shows a
peak performance of $4.18$ GFlop/s. On the GPU, counting the number of FLOPS
accurately is more difficult. We measure the performance of the \code{loop}
kernel that evaluates the equations for each pair of source and destination
particles, which in the current example is the most expensive kernel call and
takes 60\% of the total execution time. Fig.~\ref{fig:gpu_flops} shows the FLOPS
for the \code{loop} kernel on a Tesla T4 GPU using single and double
precision. This performance peaks at about 180 GFlop/s using single precision,
and about $80$ GFlop/s using double precision. We note that at this point our
GPU code still requires optimizations and that the performance will be improved
in the future.
We show the performance and scale-up of PySPH on multi-core CPUs, and GPUs by
comparing the time taken to simulate the dam break 3D problem discussed earlier
in this section with varying number of particles on different platforms. We
obtain the scale-up seen with respect to the time taken on a single-core CPU. We
use an Intel i5-7400 CPU with 4 physical cores, an NVIDIA 1050Ti GPU, and an
NVIDIA 1070Ti GPU. The speed-up obtained is shown in the
Fig.~\ref{fig:speedup}. It can be seen that on a CPU with OpenMP there is nearly
a 4 fold speed up. On the GPU with the same source code we obtain speed-up of
around 17 and 32. It is to be noted that the GPU simulations are made with
single precision.

\begin{figure}
  \centering
  \includegraphics[width=0.6\textwidth]{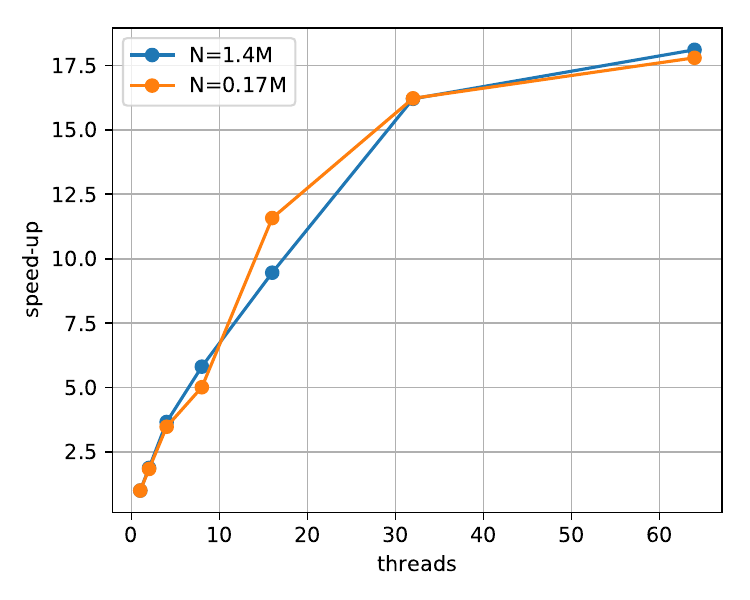}
  \caption{The OpenMP speed-up with respect to single core CPU with 0.17M and
    1.4M particles in a 3D dam-break problem.}\label{fig:omp_perf}
\end{figure}

The multi-core CPU support in PySPH is more mature than the GPU support. To
demonstrate its efficiency, we consider a 3D dam-break problem with $0.17$M
and $1.4$M particles and measure the speed-up obtained when using OpenMP.
Fig.~\ref{fig:omp_perf} shows the results as the number of threads are
increased on a machine with dual-Intel Xeon E5-2650 v3 CPUs resulting in a
total of 20 physical cores. We obtain a speed-up of around $18$ with the 20
cores for both the problem sizes. This shows that the OpenMP implementation
scales well.

\begin{figure}
  \centering
  \includegraphics[width=0.6\textwidth]{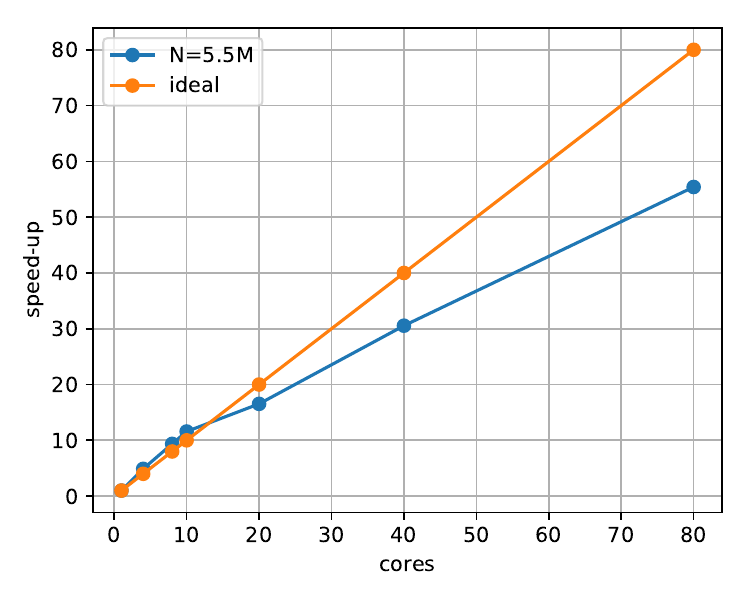}
  \caption{The MPI speed-up with respect to single core CPU with 5.5M particles
    in a 3D dam-break problem.}\label{fig:mpi_perf}
\end{figure}

We show the speed-up with MPI for the same problem in the
Fig.~\ref{fig:mpi_perf} with $5.5$M particles. We obtain a speed-up of $55$ with
80 MPI processes on a Cray cluster (with Cray OS, 2X Intel Skylake 6148 2.4 GHz
20C processors and Cray Aries with Dragonfly topology interconnect network) with
respect to a single-core simulation. This shows that PySPH performance scales
reasonably well when run in a distributed mode. We would like to add that this
aspect of PySPH can be further optimized and this is something that will be done
in the future.
\begin{table}[!ht]
\centering
\begin{tabular}{lll}
  \toprule
  & PySPH v1.0b1 & DualSPHysics v5.0 \\
  \midrule
  1-core (in secs)      & 329.55       & 146.63 \\
  OpenMP 4-cores (in secs) & 90.82        & 41.26 \\
  \bottomrule
\end{tabular}
\caption{Comparison of PySPH with DualSPHyics~\cite{crespo2015dualsphysics} on
  single-core CPU, and with OpenMP on
  4-cores.}\label{tab:comparison:dualsphysics}
\end{table}
In order to see how well PySPH performs with a well-established, fast
implementation, we compare the performance of the CPU execution with that of
DualSPHysics v5.0~\cite{crespo2015dualsphysics} on a Intel i5-7400 single and
multi-core setup. In Table~\ref{tab:comparison:dualsphysics} we note that our
performance is around twice as slow. This is because DualSPHysics
uses a customized loop to compute the neighbors and the particle accelerations
in one step. Although this is fast, it does not allow us to replace the NNPS
algorithm easily or perform iterated evaluations of equations efficiently. We
are contemplating using this approach in the future for the weakly-compressible
schemes in PySPH. Our design allows us to implement this transparently.

In this section, we have demonstrated that the automatic code generation
produces code that is easy to write but still retains good performance with
good scaling on OpenMP and MPI. The code can be executed on a GPU, however,
the performance still requires further optimization.
%

\section{Conclusions and future directions}
\label{sec:conclusions}

This article provides an overview of the design of PySPH, a powerful and
general purpose framework for smoothed particle hydrodynamics simulations. We
demonstrate how a standard problem can be solved using the framework using a
minimum of code. The overall design and underlying details of how this is
accomplished is shown. Various results demonstrating the wide applicability of
PySPH are shown.

PySPH is a unique framework for the following reasons:
\begin{itemize}
\item One can implement an entire SPH formulation in pure Python with very
  little code. This includes, SPH kernels, inter-particle interactions, and
  integrators.
\item This pure Python code can be executed without any changes on multi-core
  CPUs, GPUs, and on a cluster using MPI.
\item It implements many of the current state-of-the-art SPH schemes with over
  twenty schemes supporting different problems including weakly-compressible,
  incompressible, compressible fluid flow, elastic dynamics, DEM, etc.
\item It provides over ninety examples that can easily reused by researchers.
\item It provides many useful utilities including a built-in 3D viewer, the
  ability to share results through jupyter notebooks, and Binder.
\item Makes it easy to perform reproducible research in the area of particle
  methods.
\end{itemize}

These show that PySPH is a powerful tool that can be useful for research in
particle methods. PySPH uses object-orientation to provide powerful interfaces
that facilitate significant code reuse. Code generation is used heavily in
PySPH so as to generate hardware specific implementations on both multi-core
CPUs as well as GPUs from the same initial Python code. Finally, the use of
standard software engineering practices improve quality, and accessibility to
the software and documentation.

We plan the following improvements:
\begin{itemize}
\item Improved support for adaptive particle resolution.
\item Improved performance on GPUs and support for multiple GPUs.
\item Improved parallel performance with MPI.
\item Implementation of tree-codes to facilitate gravitational interactions,
  long-range electromagnetic forces, and the like.
\item Implementation of other meshless methods like the RKPM and MLS methods.
\end{itemize}
PySPH is already being used by some researchers across the world and we
believe that its usage will increase given its many features.

\section*{Acknowledgments}

We thank everyone who has used PySPH in their work and sent their
improvements, and reported issues. The authors would like to thank the
Department of Aerospace Engineering, Indian Institute of Technology Bombay for
the generous support as they worked on the development of PySPH. In
particular, the true academic environment of IIT Bombay has allowed the first
author the freedom to work on building tools for computational science. We
thank Prof.~Amuthan Ramabathiran, for his valuable remarks on the manuscript.
We are grateful to the anonymous reviewers whose suggestions have improved
the manuscript.


\bibliographystyle{model6-num-names}
\bibliography{references}

\end{document}